# Extending the multi-level method for the simulation of stochastic biological systems

**Christopher Lester** · **Ruth E. Baker** · **Michael B. Giles** · **Christian A. Yates**

**Abstract** The multi-level method for discrete state systems, first introduced by Anderson and Higham (2012), is a highly efficient simulation technique that can be used to elucidate statistical characteristics of biochemical reaction networks. A single point estimator is produced in a cost-effective manner by combining a number of estimators of differing accuracy in a telescoping sum, and, as such, the method has the potential to revolutionise the field of stochastic simulation.

In this paper we present several refinements of the multi-level method which render it easier to understand and implement, and also more efficient. Given the substantial and complex nature of the multi-level method, the first part of this work reviews existing literature, with the aim of providing a practical guide to the use of the multi-level method. The second part provides the means for a deft implementation of the technique, and concludes with a discussion of a number of open problems.

## 1 Introduction

Experimental researchers such as Elowitz et al (2002), Fedoroff and Fontana (2002), Arkin et al (1998) and Barrio et al (2006) have demonstrated the stochastic nature of a range of biological phenomena. In particular, stochastic effects often affect systems characterized by low molecular populations (Székely et al, 2012), but systems with large molecular populations can also be affected under certain circumstances (Erban et al, 2009). As such, in attempting to use mathematical and computational modeling

Christopher Lester, Ruth E. Baker and Michael B. Giles
Mathematical Institute, Woodstock Road, Oxford, OX2 6GG
E-mail: lesterc@maths.ox.ac.uk

Christian A. Yates
Department of Mathematical Sciences, Claverton Down, Bath, BA2 7AY

to understand the dynamics of certain biological systems, it may be necessary to explicitly account for intrinsic noise. For example, deterministic models may provide misleading results as they are unable to account for effects such as system bistability (Székely et al, 2012), stochastic focusing (Paulsson et al, 2000) and stochastic resonance (Hou and Xin, 2003).

In this work we will focus on spatially homogeneous population-level models, which record only the numbers of each molecule type within the system. The temporal evolution of the molecular abundancies will be described by the chemical master equation (CME), which comprises a system of ordinary differential equations (ODEs). For each possible system state, the CME provides an ODE describing how the probability that the system is in this particular state changes over time. For very simple systems, a closed-form, analytic solution can be obtained (Jahnke and Huisinga, 2007), however, any complication is likely to frustrate an analytic approach. Under particular circumstances, specialized numerical approaches may be feasible (Jahnke and Huisinga, 2007, 2008; Engblom, 2009; Jahnke and Udrescu, 2010; Jahnke, 2011), but in general the high dimensionality of the problem remains a challenge and stochastic simulation is the only viable alternative. In order to understand the behavior of a particular system we generate a large number of sample paths using our stochastic simulation method of choice, and use them to calculate ensemble statistics.

The Gillespie direct method (DM) simulation method (Gillespie, 1976) is exact, in the sense that it is derived from the same fundamental hypotheses as the CME and so is rigorously equivalent to it. Several variations of Gillespie's exact DM algorithm have since been developed (Gibson and Bruck, 2000; Cao et al, 2004; McCollum et al, 2006; Li and Petzold, 2006; Anderson, 2007). We will call these exact stochastic simulation algorithms (and, for brevity, eSSAs). However, chemical reaction networks can be highly complex (Székely et al, 2012) and these eSSAs work by simulating each and every reaction sequentially. This means that even the most efficient formulation will be too slow for many practical applications. In this light, the development of efficient approximate stochastic simulation algorithms (aSSAs) that avoid the pitfalls of having to detail every single reaction is currently of great interest, and a substantial range of computational techniques have been developed in an attempt to tackle this problem (Gibson and Bruck, 2000; Gillespie, 2001; El Samad et al, 2005; Auger et al, 2006). These include various forms of the tau-leaping aSSA (Yates and Burrage, 2011; Cao et al, 2006, 2007; Gillespie, 2001; Cao et al, 2004; El Samad et al, 2005) that function by carrying out multiple reactions per step. A number of helpful survey papers provide a full account of these widely-known techniques (for example, the authors recommend Gillespie (2005) or Higham (2008)). Nonetheless, despite these advances, we remain at a stage where elucidating the behavior of many stochastic biochemical signaling networks lies beyond our reach.

This work is aimed at readers comfortable with Monte Carlo simulation: the focus will be on improving the discrete-state multi-level technique first introduced by Anderson and Higham (2012). The multi-level method provides such huge computational savings that it has the potential to significantly alter the field of stochastic simulation. Given the substantial and complex nature of this approach, we first review the multi-level method, which uses a clever combination of simulation methods of varying degrees of accuracy to estimate the system statistics of interest. The idea is to compute many (cheap) sample paths at low accuracy and correct the statistics generated from them using fewer (expensive) sample paths at high accuracy. Thereafter, we consider a number of refinements to the multi-level method that render it much easier to understand and implement, and also more efficient. We also provide sample code in MATLAB and C++ in order to facilitate rapid and straightforward implementation.

1.1 Outline

In Section 2 we succinctly provide background material which allows us, in Section 3 to introduce the multi-level method of Anderson and Higham (2012). We provide a practical approach to implementing this efficient simulation technique and present a number of novel refinements to it. In Section 4 we present an in-depth discussion of methods for choosing the multi-level parameters, and in Section 5 we show results from a second example system. Whilst the first example allows us to explore the results of Anderson and Higham (2012) directly, this second example exhibits different dynamic behaviour and therefore presents different simulation challenges. A third example is presented in Section 6; it demonstrates the effectiveness of the multi-level technique on relatively complicated reaction networks. Finally, we conclude, in Section 7, with a brief discussion. All results shown here were generated in either MATLAB or C++ using a desktop computer, which was equipped with a 4.2 GHz AMD FX(tm)-4350 processor, and eight gigabytes of RAM.

## 2 The chemical master equation setting

We consider a biochemical network comprising $N$ species, $S_1,\ldots,S_N$, that may each be involved in $M$ possible interactions, $R_1,\ldots,R_M$, which are referred to as reaction channels. For the purpose of this discussion, we will ignore spatial effects. This is a reasonable assumption if the molecules are well-stirred, or, in other words, evenly distributed throughout a given volume. The population size of $S_i$ is known as its copy number and is denoted by $X_i(t)$ at time $t$, $t \geq 0$. The state vector is then defined as

$$X(t) := \begin{bmatrix} X_1(t) \\ \vdots \\ X_N(t) \end{bmatrix}. \qquad (1)$$

With each reaction channel, $R_j$, we associate two quantities. The first is the stoichiometric or state-change vector,

$$\nu_j = \begin{bmatrix} \nu_{1j} \\ \vdots \\ \nu_{Nj} \end{bmatrix}, \qquad (2)$$

where $\nu_{ij}$ is the change in the copy number of $S_i$ caused by reaction $R_j$ taking place. Thus if the system is in state $X$ and reaction $R_j$ happens, the system jumps to state $X + \nu_j$. The second quantity is the propensity function, $a_j$. This represents the rate at which a reaction takes place. Formally, for small $dt$, and based on a condition of $X(t) = x$, we define $a_j(x)$ as follows:

- the probability that reaction $R_j$ happens exactly once during the infinitesimal interval $[t, t + dt)$ is $a_j(x)dt + o(dt)$;
- the probability of more than one reaction $R_j$ during this interval is $o(dt)$.

| Reaction | Example | Propensity |
|---|---|---|
| Zero-order | $\emptyset \xrightarrow{c_1} S_1$ | $c_1$ |
| First-order | $S_1 \xrightarrow{c_2} S_2$ | $c_2 \cdot X_1$ |
| Second-order | $S_1 + S_2 \xrightarrow{c_3} S_3$ | $c_3 \cdot X_1 \cdot X_2$ |
| Homo-dimer formation | $S_1 + S_1 \xrightarrow{c_4} S_4$ | $c_4 \cdot X_1 \cdot (X_1 - 1)$ |

**Table 1** Sample reaction propensities for a stochastic system. Note for the propensity of homo-dimer formation, we have adopted the common practice of absorbing the multiplier $1/2$ into $c_4$.

Since we have assumed that the system is well-stirred, it seems reasonable for the propensity function $a_j$ of reaction $R_j$ to be proportional to the number of possible combinations of reactant molecules in the system. For example, we expect that a reaction of the type $S_1 \to S_2$, where one $S_1$ molecule becomes one $S_2$ molecule, will broadly occur at a rate proportional to the abundance of $S_1$. In second-order reactions, such as $S_1 + S_2 \to S_3$, the rate should be proportional to the abundance of pairs of $(S_1, S_2)$ molecules. Full details are given in Table 1.

Our approach to understanding the dynamics of the system comes from considering how the probability that the system is in a particular state changes through time. Define

$$\mathbb{P}(x, t \mid x_0, t_0) := \mathbb{P}\left[X(t) = x, \text{ given } X(t_0) = x_0\right]. \tag{3}$$

By considering the possible changes in species numbers brought about by a single reaction taking place, it is possible to arrive at the aforementioned CME (Gillespie, 2005):

$$\frac{d\mathbb{P}(x, t \mid x_0, t_0)}{dt} = \sum_{j=1}^{M} [\mathbb{P}(x - \nu_j, t \mid x_0, t_0) \cdot a_j(x - \nu_j) - \mathbb{P}(x, t \mid x_0, t_0) \cdot a_j(x)]. \tag{4}$$

2.1 The Kurtz Representation

Thus far we have used the propensity function and stoichiometric vector of each reaction channel to construct a CME. For completeness we also describe the Kurtz (1980) representation: this alternative method uses a construction of an inhomogeneous Poisson process for each reaction channel to represent the system. This approach provides a useful analytical tool which has been widely used in the literature.

In order to generalize homogeneous Poisson processes, we follow Kurtz (1980): suppose we have a homogeneous Poisson process of fixed rate $\lambda$ labeled as $Y_\lambda$. Then further suppose we have a Poisson process of unit rate, $Y_1$. As Poisson processes count the number of 'arrivals' over time, they can be compared by considering the distribution of the number of arrivals by some time $t$. If $Y_\lambda(t)$ and $Y_1(t)$ represent the number of arrivals by a time $t$ in the two processes, then there is an equality in distribution, that is $Y_\lambda(t) \sim Y_1(\lambda t)$. It is therefore possible to rescale time to transform a unit rate Poisson process to one of arbitrary (but known) rate.

We generalize the process by letting $\lambda$ at time $t$ be a function of the system history over the time interval $[0,t)$, as well as the system time. The number of arrivals by time $t$ is given by:

$$\overline{Y}(t) = Y_1\left(\int_0^t \lambda(t', \{\overline{Y}(s) : s < t'\}) dt'\right), \tag{5}$$

where $\lambda(t', \{\overline{Y}(s) : s < t'\})$ emphasizes that $\lambda$ is a function of the particular path the process is taking.

For the case of the stochastic reaction network, it can be shown that each reaction channel corresponds to an inhomogeneous Poisson process of rate $a_j(X(t'-))$ (Anderson et al, 2011). The $X(t'-)$ term is used to indicate the population 'just before' time $t'$. Incidentally, this ensures the process is Markovian. For reaction channel $R_j$ we let

$$Y_j\left(\int_0^t a_j(X(t'-)) dt'\right)$$

represent the number of reactions fired by time $t$. To represent the evolution of the entire network dynamics we take the appropriate sum over all reaction channels:

$$X(t) = X(0) + \sum_{j=1}^M Y_j\left(\int_0^t a_j(X(t'-)) dt'\right) \cdot \nu_j. \tag{6}$$

2.2 Example

As an example system with which to work, we consider a model of gene transcription and translation, as introduced by Anderson and Higham (2012):

$$R_1 : G \xrightarrow{25} G+M, \quad R_2 : M \xrightarrow{1000} M+P, \quad R_3 : P+P \xrightarrow{0.001} D, \tag{7}$$

$$R_4 : M \xrightarrow{0.1} \emptyset, \quad R_5 : P \xrightarrow{1} \emptyset.$$

A molecule of mRNA ($M$) is transcribed from a single gene ($G$). This mRNA molecule is then used in the translation of a protein molecule ($P$). Two protein molecules may combine to produce stable homodimers ($D$), whilst both the mRNA and protein decay linearly. We assume that the system contains a single copy of the gene throughout, and that initially there are no copies of $M$, $P$ or $D$. We write the numbers of mRNA, protein and dimer molecules at time $t$, respectively, as $X^T(t) = (X_1(t), X_2(t), X_3(t))^T$ and consequently the initial condition can be expressed as $X^T(0) = (0,0,0)^T$. The system consists of five reaction channels, as labeled in equation (7), and the corresponding stoichiometric matrix is

$$v = \begin{bmatrix} 1 & 0 & 0 & -1 & 0 \\ 0 & 1 & -2 & 0 & -1 \\ 0 & 0 & 1 & 0 & 0 \end{bmatrix}. \tag{8}$$

Due to the presence of a bimolecular reaction in this system, it is not possible to write down an analytic solution of the CME. Moreover, due to the high dimensionality of the system state space, numerical approximation of the CME is also impossible using currently available approaches (Higham, 2008; Jahnke and Huisinga, 2007). In this light, the system must be explored using a suitable eSSA or aSSA.

2.3 An exact stochastic simulation algorithm

The simplest, and perhaps most widely used eSSA for generating sample paths is Gillespie's DM (Gillespie, 1977). Suppose a reaction system has state vector $X(t_0)$ at time $t_0$, and that we wish to generate a sample path until a terminal time $T$. The DM algorithm is as follows:

1. set $X := X(t_0)$ and $t := t_0$;
2. calculate the propensity function, $a_j$, for each reaction channel, $R_j$, $j = 1, \ldots, M$, based on $X(t)$, the population vector at time $t$. Calculate the total propensity $a_0 := \sum_{j=1}^{M} a_j$;
3. generate $\Delta$, a random exponential variate with parameter $a_0$. This can be achieved by generating $r_1$ uniformly on $(0,1)$ and then setting $\Delta := (-1/a_0)\log(r_1)$. The next reaction will take place at $t + \Delta$, unless $t + \Delta > T$, in which case terminate the algorithm;
4. choose a reaction, $R_k$, to happen so that each reaction, $R_j$, $j = 1, \ldots, M$ has probability $a_j/a_0$ of being chosen. Do this, for example, by generating $r_2$ uniformly on $(0,1)$ and determining the minimal $k$ such that $\sum_{j=1}^{k} a_j > a_0 \times r_2$;
5. set $X(t+\Delta) := X(t) + v_k$ and $t := t + \Delta$ to implement reaction $R_k$ at time $t + \Delta$;
6. return to step 2.

If this algorithm is used to generate $n$ sample paths, they can be used to estimate the mean copy number of a species at a time $T$. We estimate this quantity, $\mathbb{E}[X_i(T)]$, by taking

$$\mathbb{E}[X_i(T)] \approx \frac{1}{n} \sum_{r=1}^{n} X_i^{(r)}(T), \tag{9}$$

where the copy number of species $i$ at time $t$ in path $r$ is represented by $X_i^{(r)}(t)$. This is an example of a Monte Carlo estimator, and, as such, our estimate contains a statistical error. This arises as we have studied only a subset of possible systems paths, and are therefore somewhat uncertain as to our estimate. More precisely, if the variance of our $n$ sample points is $\sigma^2$, then the estimator variance is $\sigma^2/n$. This can be used to construct a confidence interval to characterize the statistical error.

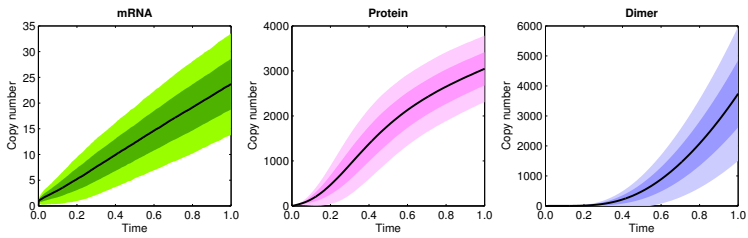

**Fig. 1** Evolution of $X^T(t) = (X_1(t), X_2(t), X_3(t))^T$ up until terminal time $T = 1$. The solid black lines show the mean species numbers and the colored bands one and two standard deviations from the mean.

| Species | Mean |
|---|---|
| $\mathbb{E}[X_1]$ | $23.79 \pm 0.004$ |
| $\mathbb{E}[X_2]$ | $3052.9 \pm 0.33$ |
| $\mathbb{E}[X_3]$ | $3714.0 \pm 0.99$ |

**Table 2** Estimated populations of system (7) at time $T = 1$, as determined by the DM, using 4,800,000 paths. 95% confidence intervals have been constructed; these are indicated with the '$\pm$' terms.

## 2.4 Example

To illustrate use of the Gillespie DM we return to the example of Section 2.2, with the aim of estimating the dimer population, $X_3(t)$, at terminal time $T = 1$. Figure 1 shows evolution of $X^T(t) = (X_1(t), X_2(t), X_3(t))^T$ up until this terminal time. The solid black lines show the mean species numbers and the colored bands one and two standard deviations from the mean.

To compute $\mathbb{E}[X_3(1)]$ to within a single dimer with 95% confidence requires the generation of approximately 4,800,000 sample paths. Using an optimized Gillespie DM algorithm, this calculation took a little over two hours (7,650 seconds) with code written in `C++`, when run on our AMD desktop computer. With code written in `MATLAB`, the same computation took approximately six hours (21,472 seconds). Table 2 shows the approximate molecular populations at the terminal time. Although our optimized DM algorithm performs adequately, the fact remains that for this system, and many others, eSSAs which simulate each reaction explicitly are prohibitively costly. In this light, a number of aSSAs have been proposed. We discuss one such aSSA, the so-called tau-leaping method, here.

## 2.5 Tau-leaping algorithm

As previously illustrated, constraints on computing resources often limit the applicability of the Gillespie DM, and other eSSAs, as they simulate each and every reaction individually. The large costs in doing so come from two main sources: first is the computational overheads in generating the large quantity of random numbers required by the algorithm; and second is the search time involved in determining which reaction occurs. One possible solution to the problem can be found in using approximate algorithms that enable one to fire a number of reactions within a single time step.

The tau-leaping method, first proposed by Gillespie (2001), generates approximate sample paths by taking steps, of length $\tau$, through time and firing several reactions during each time step. In this way it 'leaps' over several reactions at a time. If the system is in state $X$ and a time step of $\tau$ is to be performed, let $K_j(\tau, X)$ represent the number of times that reaction channel $R_j$ fires within that time step. The key, time-saving assumption of the tau-leaping method is that all reaction rates are assumed to remain constant over each time step of length $\tau$. This means that $K_j(\tau, X)$ is Poisson distributed, i.e. $K_j \sim \text{Poisson}(a_j(X(t)) \cdot \tau)$. To see this, we start from the Kurtz perspective of Section 2.1 and write

$$X(t+\tau) = X(t) + \sum_{j=1}^{M} Y_j \left( \int_t^{t+\tau} a_j(X(t'-)) \mathrm{d}t' \right) \cdot \nu_j.$$

When $t' \in [t, t+\tau)$, by assumption $a_j(X(t'-)) = a_j(X(t))$, and so we immediately deduce that $\int_t^{t+\tau} a_j(X(t'-))\mathrm{d}t' = a_j(X(t)) \cdot \tau$ as required.

The algorithm proceeds at each time step by generating Poisson variates with the correct parameter for each of the $K_j$, and then updating each molecular species and propensity function simultaneously. In general $\tau$ can be chosen adaptively as a function of the copy numbers but, for simplicity, we implement this technique by considering a fixed choice of $\tau$ throughout. This is the tau-leaping algorithm (Gillespie, 2001):

1. set $X := X(t_0)$ and $t := t_0$;
2. if $t + \tau \leq T$, calculate the propensity function, $a_j$, for each reaction channel, $R_j$, $j = 1, \ldots, M$, based on $X(t)$, the population vector at time $t$. Otherwise, exit the algorithm;
3. generate Poisson random variates, $p_j$, as sample values of $K_j(\tau, X)$, $j = 1, \ldots, M$;
4. set $X := X + \sum_{j=1}^{M} p_j \nu_j$ and $t := t + \tau$;
5. return to step 2.

As the tau-leaping algorithm produces sample paths that do not fully conform with the dynamics of the CME, any estimators calculated using the tau-leaping method may be biased: the expected difference between the tau-leaping estimate and the true value of the point estimate is non-zero. The level of bias in a tau-leaping estimate depends on the value $\tau$ takes, with lower values of $\tau$ corresponding to a lower bias. As $\tau \downarrow 0$, the estimator bias tends to zero (Anderson et al, 2011). However, the algorithm takes $\mathcal{O}(1/\tau)$ units of time to generate a path, and so generating a tau-leaping estimator with low bias requires a high level of computational resources.

## 3 Discrete-state multi-level simulation

The multi-level method of Anderson and Higham (2012) divides the work done in calculating a point estimate of the system into parts, known as levels, in an effort to increase computational efficiency. Much of the approach taken by Anderson and Higham (2012) emulates that taken by Giles (2008), in the context of numerically approximating the solutions of stochastic differential equations. Suppose we wish to estimate the expected value of $X_i(T)$, the population of the $i$-th species at time $T$. On the base level (level 0), a tau-leaping aSSA with a large value of $\tau$ (which we denote $\tau_0$) is used to generate a large number ($n_0$) of sample paths of the system. The resulting point estimate is

$$Q_0 := \mathbb{E}\left[Z_{\tau_0}\right] \approx \frac{1}{n_0} \sum_{r=1}^{n_0} Z_{\tau_0}^{(r)}(T), \qquad (10)$$

where $Z_{\tau_0}^{(r)}(t)$ is the copy number of species $i$ at time $t$ in path $r$ generated using the tau-leaping method with time step $\tau_0$, and $n_\ell$ is the number of paths generated on level $\ell$. As $\tau_0$ is large, this estimate is calculated cheaply (recall that tau-leaping takes $\mathcal{O}(1/\tau_0)$ units of time to generate each sample path), with the downside being that it is of considerable bias.

The goal with the next level (level 1) is to introduce a correction term that begins to reduce this bias. In essence, in order to compute this correction term, two sets of $n_1$ sample paths are calculated. One set comes from the tau-leaping aSSA with the same value of $\tau$ as on the base level ($\tau_0$). The other set comes from the same tau-leaping aSSA, but with a smaller value of $\tau$ (which we denote $\tau_1$). The correction term is the difference between the point estimates calculated from each set of sample paths:

$$Q_1 := \mathbb{E}\left[Z_{\tau_1} - Z_{\tau_0}\right] \approx \frac{1}{n_1} \sum_{r=1}^{n_1} \left[Z_{\tau_1}^{(r)}(T) - Z_{\tau_0}^{(r)}(T)\right].$$

Adding this correction term to the estimator calculated on the base level gives an overall more accurate estimator. This can be seen by noting that $Q_0 + Q_1 = \mathbb{E}\left[Z_{\tau_0}\right] + \mathbb{E}\left[Z_{\tau_1} - Z_{\tau_0}\right] = \mathbb{E}\left[Z_{\tau_1}\right]$, so that the sum of the two estimators has a bias equivalent to that of the tau-leaping method with $\tau = \tau_1$. The key to the efficiency of the multi-level method is to generate the two sets of sample paths,

$$\left\{Z_{\tau_1}^{(r)}(T), Z_{\tau_0}^{(r)}(T) : r = 1, \ldots, n_1\right\},$$

in a clever way, so that the variance in their difference is minimised. If the variance in their difference is denoted as $V_\ell$, then the estimator variance is given as $V_\ell/n_\ell$. A lower sample variance will mean that fewer sample paths need to be generated to achieve the same estimator variance.

On the level 2, this process is repeated to give a second correction term. Two sets of $n_2$ sample paths are generated, one set has $\tau = \tau_1$, and the second has $\tau = \tau_2 < \tau_1$. Again, the correction term is the estimator of their difference,

$$Q_2 := \mathbb{E}\left[Z_{\tau_2} - Z_{\tau_1}\right] \approx \frac{1}{n_2} \sum_{r=1}^{n_2} \left[Z_{\tau_2}^{(r)}(T) - Z_{\tau_1}^{(r)}(T)\right],$$

and it is added to the combined estimator from the level 0 and level 1 to give $Q_0 + Q_1 + Q_2 = \mathbb{E}\left[Z_{\tau_2}\right]$. Carrying on in this way, the multi-level method forms a telescoping sum,

$$\mathbb{E}\left[Z_{\tau_L}\right] = \mathbb{E}\left[Z_{\tau_0}\right] + \sum_{\ell=1}^{L} \mathbb{E}\left[Z_{\tau_\ell} - Z_{\tau_{\ell-1}}\right] = \sum_{\ell=0}^{L} Q_\ell.$$

With the addition of each subsequent level the bias of the estimator is reduced further, until a desired level of accuracy is reached.

Finally, and optionally, by generating two sets of $n_{L+1}$ sample paths, one set using an eSSA and the other using tau-leaping with $\tau = \tau_L$, we can efficiently compute a final correction term,

$$Q^*_{L+1} = \mathbb{E}\left[X_i - Z_{\tau_L}\right] \approx \frac{1}{n_{L+1}} \sum_{r=1}^{n_{L+1}} \left[X_i^{(r)}(T) - Z_{\tau_L}^{(r)}(T)\right],$$

which can be added to the telescoping sum in order to make the estimator unbiased and hence give

$$\mathbb{E}\left[X_i\right] = \mathbb{E}\left[Z_{\tau_0}\right] + \sum_{\ell=1}^{L} \mathbb{E}\left[Z_{\tau_\ell} - Z_{\tau_{\ell-1}}\right] + \mathbb{E}\left[X_i - Z_{\tau_L}\right] = \sum_{\ell=0}^{L} Q_\ell + Q^*_{L+1}. \quad (11)$$

Importantly, it turns out that the total time taken to generate the sets of sample paths for the base level, $Q_0$, and each of the correction terms, $Q_\ell$ for $\ell = 1, \ldots, L$, and $Q^*_{L+1}$, can be less than that taken to estimate $\mathbb{E}[X_i(T)]$ using an eSSA. In Section 3.3.1 we describe the bespoke simulation method used to provide samples for the correction terms which is responsible for this time-saving.

The biased estimator,

$$\mathbb{Q}_b := \sum_{\ell=0}^{L} Q_\ell, \quad (12)$$

is influenced by two distinct types of error: a *statistical error*, and a *bias*. Taking into account the choice of $\tau_0$, the bias is controlled by having sufficiently many correction terms within the sum. This will be discussed in detail later in this guide.

The statistical error is controlled by bounding the associated estimator variance, $\mathbb{V}_b$, by a parameter $\varepsilon^2$. Each of the estimators, $Q_\ell$, which make up $\mathbb{Q}_b$ has an estimator variance associated with it: if the estimate on level $\ell$ has sample variance $V_\ell$, and was calculated using $n_\ell$ samples, then the estimator variance is $V_\ell / n_\ell$. Since each level is estimated in an independent fashion, $\mathbb{V}_b = \sum_{\ell=0}^{L} V_\ell / n_\ell$. Therefore, if $n_\ell$ is sufficiently large, we will have ensured that the overall estimator variance is below a given threshold. That is, $\mathbb{V}_b < \varepsilon^2$. The unbiased estimator

$$\mathbb{Q}_u := \sum_{\ell=0}^{L} Q_\ell + Q^*_{L+1}, \quad (13)$$

also suffers from a statistical error, and it can be controlled in much the same way as described for the biased case.

To use the multi-level method, a number of decisions have to be made. We must consider:

- the choice of levels in the algorithm. This affects both the simulation time and the bias of $\mathbb{Q}_b$, and is determined by both $L$ and the values of $\tau_0, \tau_1, \ldots, \tau_L$;
- the values the target estimator variance, $V_\ell / n_\ell$, should take on each level, $\ell$. This ensures statistical accuracy, and also affects the simulation time;
- the choices of simulation techniques for the base level (0), the correcting levels, $1, \ldots, L$, and (if desired) the final level $L + 1$.

We will now discuss each of these choices.

### 3.1 The time step

We let $K \in \{2, 3, \dots\}$ be a scaling factor and take $\tau_\ell = \tau_{(\ell-1)}/K$ so that

$$Q_0 \equiv \mathbb{E}[Z_{\tau_0}],$$
$$Q_1 \equiv \mathbb{E}[Z_{\tau_0/K} - Z_{\tau_0}],$$
$$Q_2 \equiv \mathbb{E}[Z_{\tau_0/K^2} - Z_{\tau_0/K}],$$
$$\vdots$$
$$Q_\ell \equiv \mathbb{E}[Z_{\tau_0/K^L} - Z_{\tau_0/K^{L-1}}].$$

This means that the intervals are nested, with the same scaling factor between each, and it renders the algorithm more simple to understand and implement.

### 3.2 The estimator variance

The aim is to minimize the total expected computational time, subject to the overall estimator variance, $\mathbb{V}_b$, being sufficiently small. This is therefore a constrained optimization problem; on each level we will choose the number of sample paths, $n_\ell$. If each sample path on level $\ell$ takes $c_\ell$ units of time to generate, and the estimator of interest on level $\ell$ has population variance $V_\ell$, then we minimize the total expected computational time, subject to a suitable statistical error[1] :

$$\sum_{\ell=0}^{L} n_\ell c_\ell \quad \text{such that} \quad \sum_{\ell=0}^{L} \frac{V_\ell}{n_\ell} < \varepsilon^2. \tag{14}$$

In this case, $\varepsilon^2$ controls the estimator variance of our combined estimator. We performed the required optimization using Lagrange multipliers. We seek a $\lambda \in \mathbb{R}$ such that

$$\frac{\partial}{\partial n_\ell} \left[ \sum_{m=0}^{L} c_m n_m + \lambda \sum_{m=0}^{L} \frac{V_m}{n_m} \right] = 0 \quad \text{for} \quad \ell = 0, 1, \dots, L.$$

This implies $n_\ell = \sqrt{\lambda \cdot V_\ell / c_\ell}$. As we require $\sum_{m=0}^{L} V_m / n_m < \varepsilon^2$, it follows that $\sqrt{\lambda} = \sum_{m=0}^{L} \sqrt{V_m \cdot c_m} / \varepsilon^2$. Therefore, each $n_\ell$ should be chosen to be

$$n_\ell = \left\{ \frac{1}{\varepsilon^2} \sum_{m=0}^{L} \sqrt{V_m \cdot c_m} \right\} \sqrt{V_\ell / c_\ell}. \tag{15}$$

Of course, this approach is helpful if the values of $V_\ell$ and $c_\ell$ are known. Whilst it may be possible to estimate $c_\ell$ as $c_\ell \approx (\tau_0/K^\ell)^{-1}$, the population variances, $V_\ell$, will often not be known analytically. However, these can be estimated using the sample variances, which have been generated from a small number of preliminary simulations.

---

[1] Note that here, and throughout the rest of this work, we implicitly include the final exact coupling level in our summations, where appropriate.

For example, Anderson and Higham (2012) generate 100 initial sample paths on each level as a basis for estimating these quantities

The time taken for each level is the product of the number of paths, $n_\ell$, and the time taken for each path on that level, $c_\ell$. Thus, the total amount of time for the multi-level simulation is

$$\frac{1}{\varepsilon^2} \left\{ \sum_{\ell=0}^{L} \sqrt{c_\ell V_\ell} \right\}^2, \qquad (16)$$

units of CPU time.

### 3.3 The estimation techniques

We now outline how to calculate the estimates for each level. These include:

- the base level, $Q_0$. This can be handled with the regular tau-leaping algorithm with time step $\tau_0$, as described in Section 2.5;
- the tau-leaping correction terms, $Q_\ell$, for $\ell \in \{1, \ldots, L\}$. This is discussed in detail below;
- the exact SSA coupled to tau-leaping correction term, $Q_\ell^*$. This is discussed in detail.

#### 3.3.1 The tau-leaping correction terms

Given our choices of $\tau_\ell$, $\ell = 1, \ldots, L$, we have that

$$Q_\ell = \mathbb{E}\left[ Z_{\tau_0/K^\ell} - Z_{\tau_0/K^{\ell-1}} \right] \approx \frac{1}{n_\ell} \sum_{r=1}^{n_\ell} \left[ Z^{(r)}_{\tau_0/K^\ell} - Z^{(r)}_{\tau_0/K^{\ell-1}} \right], \qquad (17)$$

where $Z^{(r)}_\eta$ represents the population of the $i$-th species at a time $T$ in the $r$-th sample path generated using tau-leaping with time step $\eta$ ($= \tau_0/K^\ell$ or $\tau_0/K^{\ell-1}$). The idea underlying the multi-level method is to generate sample paths to estimate (17) so that $Q_\ell$ has a low sample variance. This means that few sample paths will be required to attain a specified statistical error.

To generate the $r$-th sample value, $Z^{(r)}_{\tau_0/K^\ell} - Z^{(r)}_{\tau_0/K^{\ell-1}}$, we will need to simultaneously generate two sample paths using tau-leaping, but with different time steps. As we are constructing a Monte Carlo estimator, we require each of the sample values, $Z^{(r)}_{\tau_0/K^\ell} - Z^{(r)}_{\tau_0/K^{\ell-1}}$, to be independent of the other bracketed terms. The key point to note is that for each sample there is no need for $Z^{(r)}_{\tau_0/K^\ell}$ and $Z^{(r)}_{\tau_0/K^{\ell-1}}$ to be independent of one another. This is because our estimator $Q_\ell$ is not dependent on the actual copy numbers within each system, but merely their difference. By recalling that

$$\text{Var}\left[Z_{\tau_0/K^\ell} - Z_{\tau_0/K^{\ell-1}}\right] = \text{Var}\left[Z_{\tau_0/K^\ell}\right] + \text{Var}\left[Z_{\tau_0/K^{\ell-1}}\right] - 2\,\text{Cov}\left[Z_{\tau_0/K^\ell}, Z_{\tau_0/K^{\ell-1}}\right],$$

we note it is therefore permissible, and in our interests, for $Z^{(r)}_{\tau_0/K^\ell}$ and $Z^{(r)}_{\tau_0/K^{\ell-1}}$ to exhibit a strong positive correlation as this will give rise to a lower estimator variance. We achieve this positive correlation by keeping the *r*-th sample paths of the approximate processes with time steps $\tau_0/K^\ell$ and $\tau_0/K^{\ell-1}$ as similar to each other as possible. We now describe how to do this.

For the purposes of our discussion, suppose we wish to simulate a single pair of sample paths on level $\ell$ and call the sample path with time step $\tau_\ell = \tau_0/K^\ell$ the fine resolution path and that with time step $\tau_{\ell-1} = \tau_0/K^{\ell-1}$ the coarse resolution path. Since both paths have the same initial conditions, one approach to achieving strong positive correlation between the two paths is to use the same tau-leaping process to simultaneously simulate each sample path, and aim to have each reaction channel fire a similar number of times in both systems. In doing so, however, it is crucial that the two paths are distributed as they would be if generated using a standard tau-leaping method.

The thickening property of the Poisson distribution (Norris, 1998) lets this aim be realized. Suppose $\mathscr{P}_1$, $\mathscr{P}_2$, and $\mathscr{P}_3$ are independent Poisson distributions. Then, for parameters $a > 0, b > 0$,

$$\mathscr{P}_1(a+b) \sim \mathscr{P}_2(a) + \mathscr{P}_3(b), \tag{18}$$

where $\sim$ implies equality in distribution. This means that a Poisson random variate with parameter $a+b$ can be generated by generating two Poisson variates, one with parameter $a$ and the other with parameter $b$, and then adding them. In terms of our sample paths, the thickening property implies that we can use one Poisson random variate to determine how many of a particular type of reaction happen in *both* the coarse and fine resolution systems during a time step and then 'top up' any further reactions that happen in only one of the systems using further Poisson random variates.

In practice, this can be achieved be creating 'virtual reaction channels' and we reformulate each reaction channel, $R_j$, into three virtual channels. We call these virtual channels $R_j^1$, $R_j^2$ and $R_j^3$ and define them such that:

- $R_j^1$ : reactions through this channel occur in both the coarse and fine systems;
- $R_j^2$ : reaction through this channel occur only in the coarse system;
- $R_j^3$ : reactions through this channel occur only in the fine system.

We will assign propensity functions to these channels so that reactions occur at the appropriate rates in both the coarse and fine systems.

To generate coupled sample paths, we use an algorithm that steps forward with fine-resolution time steps $\tau_\ell$. We update the propensity functions of the fine resolution system at each time step, but only update the propensity functions of the coarse resolution system every $K$ steps. In other words, we only update the propensity functions of the coarse resolution system after time steps of $\tau_{\ell-1}$.

To set out the multi-level algorithm, we let $Z_c$ and $Z_f$ be the copy numbers in the coarse and fine resolution sample paths, respectively. For each reaction channel $R_j$ we define $a_j^c$ to be its propensity function in the coarse resolution system and similarly for $a_j^f$. We will assume that $T/\tau_0$ is an integer, so that on each level we always take an integer number of time steps. With scaling factor $K$ we then have $\tau_\ell = \tau_0/K^\ell$. The algorithm then proceeds as follows:

1. set $Z_c := Z_c(t_0)$, $Z_f := Z_f(t_0)$ and $t := t_0$;
2. for $\alpha = 1$ to $T/\tau_0$:
   (a) calculate the propensity function $a_j^c$ for each reaction channel $R_j$, $j = 1, \ldots, M$;
   (b) for $\beta = 1$ to $K$:
      i. calculate the propensity function $a_j^f$ for each reaction channel $R_j$, $j = 1, \ldots, M$;
      ii. define the propensity functions of the three virtual channels as
      
      $$\begin{aligned} b_j^1 &= \min\{a_j^f, a_j^c\}, \\ b_j^2 &= a_j^c - b_j^1, \\ b_j^3 &= a_j^f - b_j^1; \end{aligned} \qquad (19)$$
      
      iii. for each of the virtual reaction channels, $r = 1, 2, 3$, generate Poisson random variates, $Y_j^r$, with parameters $b_j^r \cdot \tau_\ell$ and set
      
      $$\begin{aligned} Z_c &:= Z_c + \sum_{j=1}^M (Y_j^1 + Y_j^2)\nu_j, \\ Z_f &:= Z_f + \sum_{j=1}^M (Y_j^1 + Y_j^3)\nu_j; \end{aligned} \qquad (20)$$
      
   otherwise terminate;
3. return to step 2.

In step 2(b), $K$ time steps of duration $\tau_\ell$ are carried out, to give a time step of length $K\tau_\ell = \tau_{\ell-1}$ in total. Within each of these steps, $b_j^1$ is the reaction propensity of the virtual reaction channel $R_j^1$, $b_j^2$ is that of $R_j^2$, and $b_j^3$ that of $R_j^3$. We note that by equation (18), with a time step of $\tau_\ell = \tau_0/K^\ell$,

$$\begin{aligned} \mathscr{P}(a_j^c \cdot \tau_\ell) &\sim \mathscr{P}(b_j^1 \cdot \tau_\ell) + \mathscr{P}(b_j^2 \cdot \tau_\ell), \\ \mathscr{P}(a_j^f \cdot \tau_\ell) &\sim \mathscr{P}(b_j^1 \cdot \tau_\ell) + \mathscr{P}(b_j^3 \cdot \tau_\ell), \end{aligned} \qquad (21)$$

so that each sample paths is updated using the correct propensity functions. In Figure 2 we illustrate how the time steps are arranged on the time axis. We have shown a coarse time step of $\tau_{\ell-1} = 1/3$, and a fine time step of $\tau_\ell = 1/9$. In this case our scaling factor, $K = 3$, and so we have three steps of the fine process for every step of the coarse process.

Using the same Poisson random variates, $Y_j^1$, $j = 1, \ldots M$, to update both the coarse and fine system populations in (20) is crucial to the success of the method, and has the effect of introducing a strong path-wise correlation between the coarse and fine resolution sample paths. The premise is as follows: if the state vectors $Z_c(t)$ and $Z_f(t)$ show similar populations for each species, then we expect $a_j^c$ and $a_j^f$ to be similar for all $j$, as these are continuous functions of the underlying populations. If this is the case then the $b_j^r$, as defined in (19), are such that for all $j$, $b_j^1 \gg \max\{b_j^2, b_j^3\}$. Over the fine time step $\tau_\ell$, we can share the randomness between the coarse and fine system by using a single Poisson random number, generated with parameter $b_j^1 \cdot \tau_\ell$, to introduce $Y_j^1$ reactions in channel $R_j$ into both systems. To ensure compliance with the tau-leaping algorithm, we introduce a further $Y_j^2$ and $Y_j^3$ $j$ reactions in the coarse and fine systems, respectively. Note that at least one of $Y_j^2$ and $Y_j^3$ will be zero so that we 'top up' at most one of the systems. As we expect $Y_j^1$ to be significantly larger than both $Y_j^2$ and $Y_j^3$, the main part of the fluctuation is common to both systems. The result is that the state vectors in both systems remain comparable. The argument then repeats itself for each time step, and the population differences between equivalent species at the terminal time is therefore likely to be small and strongly correlated.

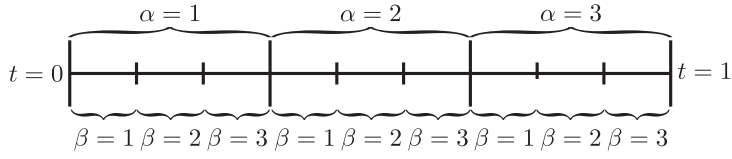

**Fig. 2** A diagrammatic representation of the steps in the algorithm, shown on a time axis, from time $t = 0$ to $t = 1$. The vertical lines represent the discretization of time.

### 3.3.2 The exact coupling

We now provide a novel technique for estimating $Q_{L+1}^*$, the final correction term that is needed to produce an unbiased point estimator, $\mathbb{Q}_u$. $Q_{L+1}^*$ is the expected difference between the point estimator generated from a tau-leaping approximation with $\tau_L = \tau_0/K^L$ and that generated using the DM eSSA. The benefit of including this final correction term into the multi-level estimator is that it allows us to produce an overall unbiased estimator, and therefore provides an output of equivalent accuracy to that of the DM. Recall that

$$Q_{L+1}^* = \mathbb{E}\left[X - Z_{\tau_0/K^L}\right] \approx \frac{1}{n_{L+1}} \sum_{r=1}^{n_{L+1}} \left[X^{(r)} - Z_{\tau_L}^{(r)}\right],$$

where $X^{(r)}$ and $Z_{\tau_L}^{(r)}$ represent the copy numbers of the species of interest at time $T$ in the $r$-th sample paths, generated by the DM, and tau-leaping aSSA with time step $\tau_L$, respectively. As for levels $\ell = 1, \ldots, L$, we aim to correlate the sample paths $X^{(r)}$ and $Z_{\tau_L}^{(r)}$ in order to reduce the variance in $Q_{L+1}^*$.

The difficulty in coupling the two sample paths arises because the tau-leaping system has its reaction propensities updated after a fixed period of time, not after a fixed number of reactions. The DM is not equipped to provide sample paths which exhibit this non-Markovian behavior. We thus have to deal with what Anderson (2007) describes as a time-inhomogeneous Poisson process. One approach to handling this situation is to use a form of the modified next reaction method (MNRM) (Anderson, 2007) to simulate the required sample paths (Anderson and Higham, 2012). We do not explore this approach further, but rather present our own simpler and more efficient method. However, our technique is mathematically equivalent to the MNRM, and therefore generates equivalent statistics. In particular, we preserve the same mean and variance. We demonstrate the improvements in simulation time in comparison to the algorithm of Anderson and Higham (2012) in Section 3.4.1.

In order to couple the same paths, we reformulate the tau-leaping algorithm so that it can be implemented in the same way as the DM, one reaction at a time. This is possible for the following reasons. Firstly, we can express a (homogeneous) Poisson process in terms of a number of exponential random variables: a Poisson process with rate $\lambda$ has inter-arrival event times exponentially distributed with rate parameter $\lambda$ (Norris, 1998). Instead of using a single Poisson random number to decide how many events occur in our Poisson process over a time interval $\tau$, we can simulate a number of inter-arrival times, and work out how many fit into the time interval $\tau$. This then gives the number of reaction events, and will be distributed in the same way as the Poisson random variable. Secondly, we can extend these results to account for parameters which are step functions in time.

We then implement the tau-leaping aSSA one reaction at a time by creating random variates that specify the time between reaction within the time step of length $\tau_L$. We will therefore simulate a number of reaction events - however, unlike the DM, the parameters (reaction propensities) are not immediately updated. We can therefore think of the generation of exponential random variables as modeling delayed reactions. This means that if we initialize a reaction at time $t \in [n\tau_L, (n+1)\tau_L)$ where $n \in \mathbb{N}$, say, we only alter the species numbers at the later time $t = (n+1)\tau_L$, so that $(n+1)\tau_L - t$ is the delay. In line with the regular tau-leaping algorithm, the propensity functions are only updated at the end of each time step. This means that our 'delayed tau-leaping' algorithm is equivalent to the regular tau-leaping algorithm. Crucially, it resembles the same structure as the DM.

Since, in this work, we are only interested in the state vector at a terminal time, we can simplify our method by thinking of delayed reactions as having an immediate effect on the population, but a deferred effect on the propensity functions. This has the effect of simplifying the internal dynamics of the algorithm, without altering our final estimate.

Let $X(t)$ represent the state vector in the exact sample path at time $t$, and similarly let $Z(t)$ represent the state vector in the tau-leaping sample path. Let $a_j$ represent the reaction propensity of reaction channel $R_j$ in the exact system, and $b_j$ represent the reaction propensity of the same channel in the tau-leaping system. Following the approach outlined previously for coupling coarse and fine tau-leaping paths, we will think of each reaction channel, $R_j$, as having three virtual channels, which we call $R_j^1$, $R_j^2$ and $R_j^3$ and define as follows:

- $R_j^1$ : this channel causes reaction $j$ to happen in both systems, and has propensity function $c_j = \min\{a_j, b_j\}$;
- $R_j^2$ : this channel causes reaction $j$ to happen in the exact system only, and has propensity function $a_j - c_j$;
- $R_j^3$ : this channel causes reaction $j$ to happen in the tau-leaping system only, and has propensity function $b_j - c_j$.

The purpose of this coupling is to share random fluctuations between the two paths. Reaction $R_j^1$ will occur far more often than $R_j^2$ and $R_j^3$, and hence the populations of both systems will remain similar. Importantly, this approach is consistent with the requirements of the system dynamics for the following reasons.

Note that the overall reaction propensity of these three virtual reactions is $d_j = \max\{a_j, b_j\}$. Moreover, if $a_j > b_j$ (therefore $c_j = \min\{a_j, b_j\} = b_j$ and $d_j = \max\{a_j, b_j\} = a_j$) we can say that if a $j$ reaction is to happen, it will definitely happen in the exact system. This is because the only way it cannot happen is if reaction channel $R_j^3$ fires, but this is impossible as $R_j^3$ has propensity of $b_j - c_j = 0$. Reaction $j$ also happens in the tau-leaping system with probability $\min\{a_j, b_j\}/\max\{a_j, b_j\} = b_j/a_j$. This follows because it is the probability of reaction channel $R_j^1$ firing, given a $j$ reaction happens. An equivalent result can be derived if $b_j > a_j$: if a $j$ reaction is to occur, it will definitely occur in the tau-leaping system, and will also occur in the exact system with probability $a_j/b_j$. Finally, if $a_j = b_j$ then, if a $j$ reaction takes place, it must do so in both systems.

We will therefore simulate the coupled system according to a two step process. The first step will be to decide which reaction, $j$, fires in at least one of the systems. The second step will be to decide whether the exact, tau-leaping or both systems are affected. We now state our algorithm for generating coupled sample paths up to a terminal time $T$:

1. set $X := X(t_0)$, $Z := Z(t_0)$ and $t := t_0$. Set the start time of the tau-leap time step as $t_B := t_0$, and the end time as $t_E := t_0 + \tau_L$;
2. calculate propensity function $b_j(Z(t_B))$ for each reaction channel $R_j$, $j = 1, \ldots, M$ in the tau-leaping system;
3. calculate the propensity function $a_j(X(t))$ for each reaction channel $R_j$, $j = 1, \ldots, M$ in the exact system;
4. for each reaction, $j$, calculate $d_j = \max\{a_j, b_j\}$ and the total *firing* propensity $d_0 := \sum_{j=1}^{M} d_j$;
5. generate $\Delta$, a random exponential variate with parameter $d_0$. This can be achieved by generating $r_1$ uniformly on $(0, 1)$ and then setting $\Delta := (-1/d_0) \log(r_1)$;
6. if $t + \Delta > T$, terminate the algorithm. Otherwise, if $t + \Delta \geq t_E$ update the tau-leaping propensities. Do this by setting $t := t_E$, $t_B := t_E$ and $t_E := t_E + \tau_L$, and returning to step 2. If $t + \Delta < t_E$, set $t := t + \Delta$ and continue to step 7;
7. choose a *firing* channel $j$ such that each reaction $j$ has probability $d_j/d_0$ of firing. Do this, for example, by generating $r_2$ uniformly on $(0, 1)$ and determining the minimal $j$ such that $\sum_{k=1}^{j} d_k > d_0 \times r_2$;
8. implement reaction $j$ in the system which has reaction propensity $d_j = \max\{a_j, b_j\}$, and update the population levels correspondingly;
9. sample $r_3$ uniformly from $(0, 1)$. If $r_3 < c_j/d_j$, then fire reaction $j$ in the other system also;
10. if $R_j$ has been implemented in the exact system, return to step 3. Otherwise, return to step 5.

In the context of computational efficiency, we expect our revised algorithm to differ from the original MNRM (Anderson and Higham, 2012), in two significant ways. Firstly in step 6, a number of time steps will be 'rejected' as the algorithm returns to step 2. This step is justified by the memoryless property. This means that a number of random variates will be 'wasted', with the extent of wastage depending on the value of $\tau_L$. In contrast, one of the main attractions of the MNRM is that it barely wastes any random numbers. However, if the generation of random numbers is a concern then $r_3$ in step 9 can be generated cheaply by recycling $r_2$ (Yates and Klingbeil, 2013). Secondly, in step 7, a *firing* channel is chosen, and then the affected paths are determined in step 8 and 9. The search in step 7 is amenable to optimization in much the same way as the DM (McCollum et al, 2006; Li and Petzold, 2006; Cao et al, 2004). The downside of the MNRM in this context is that a substantial amount of complex maintenance work needs to be carried out within the algorithm, and that choice of the next reaction involves an unavoidably time-consuming search to find the minimum within a matrix of 'next reaction times' on a substantially enlarged state space (as there are three virtual channels for every reaction channel). Optimization of the MNRM is somewhat less straightforward, but the method of Gibson and Bruck (2000) can potentially be adapted.

In the next section we test our multilevel algorithm with a range of different parameters. In particular, we compare our updated method with the original MNRM method of Anderson and Higham (2012) in Section 3.4.1.

## 3.4 Example

We are now in a position to implement the multi-level method on our example gene expression system, (7). We first give a detailed breakdown of the contribution to the simulation of each level using the canonical set of parameters given in (7). We then demonstrate the effects on the simulation efficiency of varying these parameters when generating both biased and unbiased estimators. In particular, we demonstrate how our new final coupling method has contributed to a significant performance improvement in the situation where an unbiased estimator is required.

Initially we chose $K = 3$, $\tau_0 = 1/9$ and $L = 5$. Using C++, we estimate the value of the mean dimer population, $\mathbb{E}[X_3(1)]$, to be $3,714.23 \pm 0.99$, with approximately 166 seconds of computation (where the error tolerances refer to a 95% confidence interval). The same calculation can be performed using MATLAB; in this case, we estimate $\mathbb{E}[X_3(1)]$ to be $3,714.55 \pm 1.06$ within 579 seconds of computation.

Compared with the 7,650 seconds taken for the DM using C++ (see Section 2.4; the equivalent figure for MATLAB is 21,472 seconds), the multi-level approach is approximately 46 times faster (equivalently, 37 times) for this example system with these canonical parameter values. In Table 3 we detail the contribution of each level of the multi-level estimator to the simulation time and the cumulative estimate of $\mathbb{E}[X_3(1)]$, when C++ has been used to produce the simulations. In this case, the final estimator, $Q^*_{L+1}$, contributes a relatively small proportion of the simulation time, which makes the calculation of the unbiased estimator a feasible option. We also show that the actual simulation times compare very well with those estimated by equation (16).

| Level | $\tau_{\ell-1}$ | $\tau_\ell$ | Estimate | Variance | Paths | Time |
|---|---|---|---|---|---|---|
| $Q_0$ | - | $3^{-2}$ | 3187.47 | $1.03 \times 10^6$ | $7.11 \times 10^6$ | 89.9s *(93.8s)* |
| $Q_1$ | $3^{-2}$ | $3^{-3}$ | 350.52 | 16287.10 | 420814 | 24.9s *(24.8s)* |
| $Q_2$ | $3^{-3}$ | $3^{-4}$ | 117.48 | 2666.80 | 114125 | 15.5s *(15.5s)* |
| $Q_3$ | $3^{-4}$ | $3^{-5}$ | 39.15 | 658.14 | 40972 | 12.0s *(12.1s)* |
| $Q_4$ | $3^{-5}$ | $3^{-6}$ | 13.00 | 196.09 | 17534 | 10.7s *(10.7s)* |
| $Q_5$ | $3^{-6}$ | $3^{-7}$ | 4.42 | 48.28 | 6406 | 7.0s *(6.9s)* |
| $Q_6^*$ | $3^{-7}$ | DM | 2.19 | 38.75 | 2870 | 5.8s *(5.8s)* |
| Total | | | $3714.23 \pm 0.99$ | | - | 165.8s |

**Table 3** The contribution from each level in producing an unbiased overall estimator, $\mathbb{Q}_u$, for $\mathbb{E}[X_3]$ in system (7) at $T = 1$. We have taken $\tau_0 = 1/9$, $K = 3$, and $L = 5$. In the time column, the true simulation time is shown, and the estimated simulation time is shown in brackets.

| | Matlab | | | C++ | | |
|---|---|---|---|---|---|---|
| time step ($\tau_L$) | Modified DM | MNRM | Saving | Modified DM | MNRM | Saving |
| | *seconds per 1000 paths* | | | *seconds per 1000 paths* | | |
| $1/3^4$ | 8.69 | 16.33 | 47% | 1.74 | 2.21 | 21% |
| $1/3^5$ | 8.77 | 16.69 | 47% | 1.75 | 2.21 | 21% |
| $1/3^6$ | 8.83 | 17.34 | 49% | 1.82 | 2.28 | 20% |
| $1/3^7$ | 9.38 | 18.74 | 50% | 1.95 | 2.41 | 19% |
| $1/3^8$ | 12.35 | 22.39 | 45% | 2.35 | 2.69 | 13% |
| $1/3^9$ | 22.10 | 32.79 | 33% | 3.45 | 3.49 | 1 % |

**Table 4** Various simulation times for calculation of $Q_{L+1}^*$ in system (7). In each case we have compared the simulation time for $n_{L+1} = 1000$ paths using the traditional MNRM and our novel Modified Direct Method.

### 3.4.1 Final coupling

In order to compare the performance of our new algorithm to estimate $Q_{L+1}^*$ with that of Anderson and Higham (2012), we produced a number of simulations to estimate $Q_{L+1}^*$, the final, bias-removing estimator using both MATLAB and C++. We implemented Anderson and Higham's method on our equipment in order to compare simulation times fairly. By considering $n_{L+1} = 1,000$ samples, in Table 4 we demonstrate that our approach reduces the simulation time in comparison to the original method of Anderson and Higham. All efforts were taken to use the optimal code for each approach. These results are demonstrated for a wide range of choices of $\tau_L$, the time step used to increment the coarse, tau-leaping paths. A reduction in simulation time is particularly noticeable when making use of MATLAB. The C++ implementation also shows a time-saving, except for very small choices of $\tau_L$. Such small values for $\tau_L$ fall outside the range that we would encounter when implementing the multi-level simulation algorithm (including increasingly small values of $\tau_L$ result in an increased overall simulation time, as there are more levels to simulate). Clearly this reduction in simulation time for $Q_{L+1}^*$ is problem-dependent and may vary widely from problem to problem. However, we have found significant reductions in simulation time for all the reaction networks we have tested and suggest that similar reductions will be possible for most systems.

| $\tau_0$ | $\tau_L$ | L+1 | K | Estimate | Duration |
|---|---|---|---|---|---|
| $2^{-3}$ | $2^{-8}$ | 6 | 2 | $3695.00 \pm 1.00$ | 121.4 s |
| $2^{-4}$ | $2^{-8}$ | 5 | 2 | $3695.95 \pm 1.01$ | 173.8 s |
| $3^{-2}$ | $3^{-5}$ | 4 | 3 | $3694.62 \pm 1.00$ | 121.6 s |
| $3^{-2}$ | $3^{-6}$ | 5 | 3 | $3707.85 \pm 0.98$ | 139.5 s |
| $3^{-2}$ | $3^{-7}$ | 6 | 3 | $3711.32 \pm 1.00$ | 146.3 s |
| $3^{-2}$ | $3^{-8}$ | 7 | 3 | $3714.13 \pm 0.99$ | 163.2 s |
| $3^{-2}$ | $3^{-9}$ | 8 | 3 | $3714.14 \pm 0.96$ | 180.3 s |
| $3^{-3}$ | $3^{-7}$ | 5 | 3 | $3711.23 \pm 0.99$ | 272.1 s |
| $3^{-4}$ | $3^{-7}$ | 6 | 3 | $3712.48 \pm 1.00$ | 586.6 s |
| $4^{-2}$ | $4^{-4}$ | 3 | 4 | $3695.59 \pm 1.00$ | 159.5 s |
| $4^{-2}$ | $4^{-5}$ | 4 | 4 | $3710.41 \pm 0.99$ | 182.1 s |

**Table 5** A range of biased estimators, $\mathbb{Q}_b$, of the terminal dimer population of $X_3(1)$ in our example gene expression system (7). $\tau_L$ refers to the time step used on the finest correction level. We show $L+1$, the total number of estimators used to generate $\mathbb{Q}_b$. The last column shows the CPU time taken; the estimated time, given by (16), is shown in italics. For each set of parameters, we used Equation (15) to determine how many simulations to perform on each level.

3.5 Exploring variation in algorithm parameters

Throughout the rest of this section, we focus on a C++ implementation of the multi-level method. The use of MATLAB will be demonstrated in the following example in Section 5. We now ask what the effect of changing $L$, the number of levels, $\tau_0$, the size of the time step on the base level, and $K$, the scaling factor, will be on the simulation time. In Tables 5 and 6 we demonstrate the effect of varying $L$ for three different values $K$ for biased and unbiased estimators respectively.

As previously noted, for our canonical parameter values, Table 3 suggests that an unbiased estimator comes at little additional cost to a biased estimator, and should, therefore, be preferred. However, for completeness in Table 5 we show the values of the biased estimators for a range of values of $L$ and $K$, as well as comparisons between the estimated and actual simulation times. We will return to this point in the discussion. These tables demonstrate the impact of a judicious choice of $\tau_0$ and $L$, but unfortunately shed little light on the optimal choice of $K$, the scaling factor. In the case of SDEs, Giles suggests that $K = 4$ may well be sensible (Giles, 2008).

We discuss the biased and unbiased cases separately. For the biased estimator, $\mathbb{Q}_b$, the choices of $\tau_0$ and $L$ determine the overall bias of the estimator. A larger value of $L$ will lead to a lower bias, but also to increased simulation time. For the unbiased estimator, $\mathbb{Q}_u$, the situation is less straightforward. Our view is that the simulation time is not particularly sensitive to the particular choice of $L$. However, we do note that the choice of $\tau_0$, the resolution on the base level, can substantially affect simulation time. We will therefore present an algorithm which provides for a reasonable choice of this input.

In both the biased and unbiased cases, the confidence intervals have not been faithfully attained: this is because we have predicted the number of paths necessary for the generation of each estimator with a specific variance based on an initial number of preliminary samples and this method has not turned out be accurate. Reasons for this are outlined later in the Discussion.

| $\tau_0$ | $\tau_L$ | L+2 | K | Estimate | Duration |
|---|---|---|---|---|---|
| $3^{-2}$ | $3^{-5}$ | 5 | 3 | $3713.97 \pm 1.00$ | 166.2 s |
| $3^{-2}$ | $3^{-6}$ | 6 | 3 | $3715.61 \pm 0.97$ | 156.7 s |
| $3^{-2}$ | $3^{-7}$ | 7 | 3 | $3714.23 \pm 0.99$ | 165.8 s |
| $3^{-2}$ | $3^{-8}$ | 8 | 3 | $3713.98 \pm 0.99$ | 179.2 s |
| $3^{-2}$ | $3^{-9}$ | 9 | 3 | $3714.47 \pm 0.97$ | 173.2 s |
| $3^{-3}$ | $3^{-6}$ | 5 | 3 | $3714.84 \pm 0.99$ | 280.8 s |
| $4^{-2}$ | $4^{-5}$ | 5 | 4 | $3713.94 \pm 1.00$ | 199.0 s |
| $4^{-2}$ | $4^{-6}$ | 6 | 4 | $3713.86 \pm 1.00$ | 199.6 s |
| $6^{-1}$ | $6^{-3}$ | 4 | 6 | $3714.09 \pm 0.99$ | 171.1 s |
| $6^{-1}$ | $6^{-4}$ | 5 | 6 | $3714.46 \pm 1.00$ | 173.6 s |

**Table 6** This table shows a range of exact estimators, $\mathbb{Q}_u$, of the terminal dimer population of $X_3(1)$ in our example gene expression system (7). $\tau_L$ refers to the time step used on the finest correction level for the biased estimator, $\mathbb{Q}_b$, before the final exact correction $Q^*_{L+1}$, has been added to remove the bias. We also show $L+2$, the number of terms contributing to the estimator $\mathbb{Q}_u$.

## 4 Method configuration

In the previous sub-section, as well as demonstrating the improved efficiency of our novel exact coupling method for the final level, we found that the choice of parameters for the multi-level method can have a significant effect on simulation time and estimator accuracy. In this section we provide a number of concrete suggestions for algorithmic choices that automatically determine suitable values for these tunable parameters. We will suggest:

A. an optimal choice of $\tau_0$, the resolution on the most inaccurate level;
B. optimal choices of $K$ and $L$, that is, the scaling factor and the number of levels to use in total;
C. the choice of whether to use a biased or an unbiased estimator.

4.1 Choice of the base level time step, $\tau_0$

It is tempting to assume that, since the multi-level method benefits from using many low quality population estimates which are simulated quickly, a large choice of $\tau_0$ would be prudent. The effect of choosing too large a value of $\tau_0$ is that, whilst the base level estimate, $Q_0$, may be calculated quickly, $Q_1$ and other correction terms will require increased computational time since more sample paths will be needed to correct the inaccurate base level estimate. The optimal value for $\tau_0$ may well depend on the particular choice of $K$, the scaling constant. For the purposes of this investigation, however, we fix the value of $K$. We will also consider the time step on the finest level, $\tau_L$, as fixed at a (unknown) value. Based on this, we choose a value for $\tau_0$ and, subsequently, $L$.

From equation (16) in Section 3.2 we recall that

$$\frac{1}{\varepsilon^2}\left\{\sum_{\ell=0}^{L}\sqrt{c_\ell V_\ell}\right\}^2,$$

units of CPU time are required to attain an estimator variance of $\varepsilon^2$. Recall also that $c_\ell$ represents the per-path simulation time, and $V_\ell$ the sample variance on a level $\ell$. To simplify notation, we introduce $k_\ell$, where

$$k_\ell := c_\ell V_\ell. \tag{22}$$

This gives an indication of the relative cost of producing simulated paths for level $\ell$. As in Section 3.2, $k_\ell$ can be estimated cheaply using a fixed (and relatively small) number of paths.

We take an iterative approach to optimizing the choice of $\tau_0$, beginning with an initial guess, and improving on it in subsequent iterations. Given a initial choice of $\tau_0$, $\tau_0^{(1)}$, we propose two candidates for an improved choice, $\tau_0^{(2)}$:

- a smaller choice, $\tau_0^{(2,1)} = \tau_0^{(1)}/K$;
- a larger choice, $\tau_0^{(2,2)} = \tau_0^{(1)}K$.

Making the reasonable assumption that there will be at least one level in addition to the 'base' level, we can calculate the difference in expected overall simulation times using $\tau_0^{(2,1)}$ or $\tau_0^{(2,2)}$. If using $\tau_0^{(2,1)}$ or $\tau_0^{(2,2)}$ results in a time saving compared with using $\tau_0^{(1)}$, we set our improved guess $\tau_0^{(2)}$ to equal the appropriate value. We can repeat this algorithm until we reach a choice of $\tau_0$ for which no further improvement is gained. This corresponds to a local minimum of the overall simulation time, and we take $\tau_0 = \tau_0^{(n)}$. If, by chance, we begin at a local maximum we follow the refinement process in both directions (both increasing and decreasing $\tau_0$).

In general, our iterative algorithm will require comparison of the computational complexity of generating an estimator with coarse base level time step $\tau_0^c$, with the computational complexity of generating an estimator with a fine base level time step $\tau_0^f = \tau_0^c/K$. The estimator for the coarse base level, given a desired level of accuracy, will be given by

$$\mathbb{Q} = \mathbb{E}\left[Z_{\tau_0^c}\right] + \mathbb{E}\left[Z_{\tau_0^c/K} - Z_{\tau_0^c}\right] + \sum_{\ell=2}^{L} \mathbb{E}\left[Z_{\tau_0^c/K^\ell} - Z_{\tau_0^c/K^{\ell-1}}\right], \tag{23}$$

and the estimator for the fine base level will be given by

$$\mathbb{Q} = \mathbb{E}\left[Z_{\tau_0^c/K}\right] + \sum_{\ell=2}^{L} \mathbb{E}\left[Z_{\tau_0^c/K^\ell} - Z_{\tau_0^c/K^{\ell-1}}\right]. \tag{24}$$

The majority of the levels are simulated for both choices of base level and, as such, will have the same relative cost, $k_\ell$. The terms that will have different relative costs will be $\mathbb{E}\left[Z_{\tau_0^c}\right]$ and $\mathbb{E}\left[Z_{\tau_0^c/K} - Z_{\tau_0^c}\right]$ on the coarse level (for which we will denote the relative costs as $k_0^c$ and $k_1^c$, respectively), and $\mathbb{E}\left[Z_{\tau_0^c/K}\right]$ on the fine level (for which we will denote the relative cost as $k_0^f$). We can use this knowledge to prove a theorem which will allow for acceptance/rejection of a proposed base level time step, using a simple comparison of these three proportionality constants.

**Proposition 1** *The configuration with the fine base level time step, $\tau_0^f = \tau_0^c/K$, should be preferred over that with coarse base level time step, $\tau_0^c$, if*

$$\sqrt{k_0^f} < \sqrt{k_0^c} + \sqrt{k_1^c}, \qquad (25)$$

*where we recall that $k_\ell$ represents the relative cost, and is given by $k_\ell = c_\ell V_\ell$.*

**Proof.** In order to see where this inequality comes from proceed as follows: without loss of generality set the variance target at $\varepsilon^2 = 1$. Then, the expected difference in simulation time between the estimator with the fine base level time step and the estimator with the coarse base level time step is given by

$$\left\{ \sum_{\ell=0}^{L^f} \sqrt{k_\ell^f} \right\}^2 - \left\{ \sum_{\ell=0}^{L^c} \sqrt{k_\ell^c} \right\}^2.$$

Using the fact that, for $i \geq 1$, $k_{i+1}^c = k_i^f$, and that $L^f + 1 = L^c$, we can rewrite this as

$$\left\{ \sqrt{k_0^f} - \sqrt{k_1^c} + \sum_{\ell=1}^{L^f+1} \sqrt{k_\ell^c} \right\}^2 - \left\{ \sqrt{k_0^c} + \sum_{\ell=1}^{L^f+1} \sqrt{k_\ell^c} \right\}^2.$$

Thus, after rearrangement, the net change in simulation time is

$$\left[ \sqrt{k_0^f} - \sqrt{k_0^c} - \sqrt{k_1^c} \right] \left\{ \sqrt{k_0^f} + \sqrt{k_0^c} + \sqrt{k_1^c} + 2 \sum_{\ell=2}^{L^f+1} \sqrt{k_\ell^c} \right\}. \qquad (26)$$

As the terms within the braced brackets are positive, we have the required condition.

*4.1.1 Example*

We again consider the gene expression system (7) and use our algorithm to choose $\tau_0$ in C++. First impose the choice of $K = 3$. If we take $\tau_0^{(1)} = 1/9$, then there are two alternatives to consider, $\tau_0^{(2,1)} = 1/27$ and $\tau_0^{(2,2)} = 1/3$. With 10,000 samples, we calculate estimates for the relevant proportionality constants and present the results in Table 7. We then use Theorem 1 to decide on the appropriate choice of $\tau_0$. The initial base level time step $\tau_0^{(1)}$ is coarse in comparison to the proposed base level time step $\tau_0^{(2,1)}$. Since we have $\sqrt{k_0^{(1)}} + \sqrt{k_1^{(1)}} = 4.5666 < 6.7624 = \sqrt{k_0^{(2,1)}}$, by Theorem 1 $\tau_0^{(2,1)} = 1/27$ is an inferior choice to $\tau_0^{(1)} = 1/9$. Similarly, as $\sqrt{k_0^{(2,2)}} + \sqrt{k_1^{(2,2)}} = 9.0220 > 3.5859 = \sqrt{k_0^{(1)}}$, Theorem 1 implies that $\tau_0^{(2,2)} = 1/3$ is also an inferior choice. We therefore take $\tau_0 = 1/9$.

Despite the fact that we have rejected $\tau_0^{(2,1)}$ and $\tau_0^{(2,2)}$ and thus selected $\tau_0^{(1)}$, it may be that choices of $\tau_0$ between $\tau_0^{(1)}$ and $\tau_0^{(2,1)}$ or between $\tau_0^{(1)}$ and $\tau_0^{(2,2)}$, (for example, $\tau_0 = 1/12$ or $\tau_0 = 1/7$, respectively) provide better performance than $\tau_0^{(1)}$. Fortunately, efficient multi-level simulation does not require that the choice of $\tau_0$ is exactly optimal, rather that particularly poor choices of $\tau_0$ are avoided. Our iterative procedure provides a mechanism by which a value of the base level time step, $\tau_0$, can be selected, given a value of $K$. A further benefit of our algorithm is that it does not require that $\tau_L$ be chosen at the outset.

| Guess | Estimates | |
|---|---|---|
| $\tau_0^{(1)} = 1/9$ | $\sqrt{k_0^{(1)}} = 3.5859,$ | $\sqrt{k_1^{(1)}} = 0.9807.$ |
| $\tau_0^{(2,1)} = 1/27$ | $\sqrt{k_0^{(2,1)}} = 6.7624,$ | $N/A.$ |
| $\tau_0^{(2,2)} = 1/3$ | $\sqrt{k_0^{(2,2)}} = 3.5576,$ | $\sqrt{k_1^{(2,2)}} = 5.4644.$ |

**Table 7** Details of the cost measure for each potential ensemble of estimators for the gene expression system (7) with different choices of $\tau_0$, the time step on the base level.

### 4.2 Choice of final accuracy in the biased estimator

As discussed previously, ensemble statistics collected with a biased system give rise to point estimates laden with both a bias, and a statistical error. It is sensible to combine these errors into a single quantity, as the source of an error may not be relevant to an end-user. If the population of species $i$, $X_i$, is estimated by the biased multi-level method as $\mathbb{Q}_b = \widehat{\theta}$ and the true expectation is given by $\mathbb{E}[X_i] = \theta$, then the mean-squared error (MSE) is

$$\text{MSE}(\widehat{\theta}) = \mathbb{E}[(\widehat{\theta} - \theta)^2] = \text{Var}(\widehat{\theta}) + (\text{Bias}(\widehat{\theta}, \theta))^2. \quad (27)$$

We can estimate the bias (Li, 2007) by noting that if $\mathbb{E}[Z_i]$ is an estimator generated using a tau-leaping algorithm with fixed time step $\tau$, then there exists a constant $C$ such that as $\tau \to 0$, $\mathbb{E}[Z_i] - \mathbb{E}[X_i] \approx C\tau$. The estimator variance can be estimated by noting that, for each level, $V_\ell \propto 1/n_\ell$. Equation (27) therefore suggests that the MSE can be controlled in two ways. Firstly, given $\tau_0$, we can take the number of levels, $L$, sufficiently large. This controls the MSE because incorporation of each additional level into the algorithm has the effect of approximately dividing the model bias by a factor of $K$. Secondly, we can increase the number of sample paths, $n_\ell$, on each level $\ell$ to decrease the variance of the biased estimator.

Suppose we are given a MSE allowance, $\varepsilon^2$, and have to ascribe a portion of this to the square of the bias, and the remainder to the variance. As a first attempt at a solution, we pre-assign a proportion, $\lambda \in (0,1)$ of the MSE allowance to the square of the bias, and leave $1-\lambda$ to the variance. Previous work (Giles, 2008) has made the simple choice $\lambda = 1/2$, that is, assigning half the MSE to the square of the bias, and the other half to the estimator variance. However, it is not clear how best to choose $\lambda$ for a particular system. We demonstrate the effects of varying $\lambda$ later in this work.

### 4.3 Towards an adaptive simulation approach

Recall that equation (16) estimates the units of CPU time required to attain an estimator variance of $(1-\lambda)\varepsilon^2$. If we add an additional level into the algorithm, this will reduce the bias, but more sample paths may be required on each level in order to reduce the variance of the combined estimator below the target of $(1-\lambda)\varepsilon^2$. Therefore, given a choice of $\lambda$, it is not clear how to best choose $L$ such that the computation is most efficient. We suggest using the following incremental approach to obtain MSE $= \varepsilon^2$, given a choice of $\lambda$:

1. initially work with a single level so that $L = 0$. Choose $\tau_0$ according to the algorithm of Section 4.1. Estimate $k_0$ and generate $n_0$ sample paths. This gives an estimator of $\mathbb{Q}_b = Q_0$ with desired statistical accuracy (an estimator variance of $(1-\lambda)\varepsilon^2$);
2. perform a bias test (using equation (29) or (30), below) on the estimator, $\mathbb{Q}_b = \sum_{\ell=0}^{L} Q_\ell$. If the bias is at most $\sqrt{\lambda}\varepsilon$, terminate the algorithm;
3. if not, introduce a new level into the system and let $L := L+1$. Estimate $k_L$ and the calculate the optimal number of sample paths for each level, $n_\ell$, $\ell = 0,\ldots,L$ according to (15);
4. generate the required number of sample paths;
5. return to step 2;

To evaluate the bias we note, for large $\ell$,

$$Q_\ell = \mathbb{E}\left[Z_{\tau_0/K^\ell} - Z_{\tau_0/K^{\ell-1}}\right] = \mathbb{E}\left[Z_{\tau_0/K^\ell} - X_i\right] - \mathbb{E}\left[Z_{\tau_0/K^{\ell-1}} - X_i\right]$$
$$\approx C\tau_0/K^\ell - C\tau_0/K^{\ell-1}$$
$$= (K-1)\mathbb{E}\left[X_i - Z_{\tau_0/K^\ell}\right].$$

Therefore we can estimate the bias as,

$$\mathbb{E}\left[X_i - Z_{\tau_0/K^\ell}\right] \approx \frac{Q_\ell}{K-1}, \tag{28}$$

and so to obtain MSE $= \varepsilon^2$, the algorithm should be terminated, at level $L$, when

$$|Q_\ell| \leq \sqrt{\lambda}(K-1)\varepsilon. \tag{29}$$

Note that to improve the reliability of this approach, one could follow Giles (2008) and perform the bias test using the two levels. The algorithm thus terminates with

$$\max\{K^{-1}|Q_{\ell-1}|, |Q_\ell|\} \leq \sqrt{\lambda}(K-1)\varepsilon. \tag{30}$$

The incremental approach outlined here also provides the opportunity to correct the errors inherited from the use of (initially poorly) approximated values for $k_\ell$. This can be done in several ways. Firstly, when the incremental algorithm is used, the estimates for each $k_\ell$ can be recalculated at step 3 as each additional level is added into the system. However, the use of updated $k_\ell$'s means that there may be a set of levels, $\Gamma$, where more sample paths have already been simulated than required by (15). This corresponds to the associated estimators, $Q_\ell$, $\ell \in \Gamma$, having estimator variances lower than required by the bound $\sum_{\ell=0}^{L} V_\ell < (1-\lambda)\varepsilon^2$. If we set

$$\varepsilon^* := (1-\lambda)\varepsilon^2 - \sum_{\ell \in \Gamma} V_\ell/n_\ell, \tag{31}$$

| Species | Sample Mean | Sample Variance |
|---|---|---|
| $S_1$ | $2047.4 \pm 0.73$ | 2022.8 |
| $S_2$ | $1897.3 \pm 0.73$ | 1995.8 |
| $S_3$ | $3027.2 \pm 1.00$ | 3808.6 |

**Table 8** Estimated populations of system (32) at time $T = 9$, as determined by the DM. 95% confidence intervals have been constructed; these are indicated with the '$\pm$' terms.

then we can still satisfy our overall variance target by achieving the variance target of $\varepsilon^*$ for the combined levels $\ell \in \{0,\ldots L\} \setminus \Gamma$. Note that if we re-calculate our target $\widehat{n}_\ell$ for $\ell \in \{0,\ldots L\} \setminus \Gamma$, we now require fewer sample paths for each level. It is therefore now possible that more sample paths than required have already been generated for some $\ell \in \{0,\ldots L\} \setminus \Gamma$. These levels can then be added to the set $\Gamma$ and $\varepsilon^*$ can be recalculated. This argument can be repeated until no more levels can be added to $\Gamma$.

Once we have ensured that the required bias limitation has been achieved, we then check whether the required statistical error has been achieved, and generate more sample paths if this is appropriate. In this way, we have a high degree of confidence that the algorithm has attained an estimator with the required error.

## 5 A second example system

In this section we consider a second, synthetic sample system:

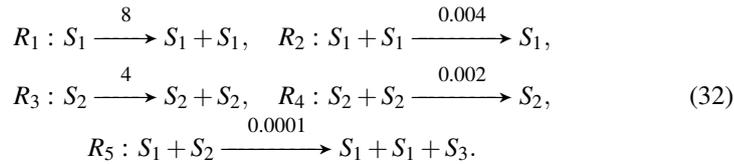

$$R_1 : S_1 \xrightarrow{8} S_1 + S_1, \quad R_2 : S_1 + S_1 \xrightarrow{0.004} S_1,$$
$$R_3 : S_2 \xrightarrow{4} S_2 + S_2, \quad R_4 : S_2 + S_2 \xrightarrow{0.002} S_2, \quad (32)$$
$$R_5 : S_1 + S_2 \xrightarrow{0.0001} S_1 + S_1 + S_3.$$

System (32) could, for example, represent a predator-prey model with $S_1$ a predator of $S_2$. $S_3$ simply counts the number of predator prey interactions. For the purposes of this discussion, we will simulate sample paths from time $t = 0$ until a terminal time $T = 9$. Throughout this section, we will make use of MATLAB to illustrate the advantages of the multi-level method. This system exhibits different dynamics to our other model system and, as such, presents different modeling challenges. Figure 3 shows evolution of $X^T(t) = (X_1(t), X_2(t), X_3(t))^T$ up until this terminal time. The solid black lines show the mean species numbers and the colored bands one and two standard deviations from the mean. To benchmark the performance of our multi-level method, we will initially estimate the mean copy numbers of $S_1$, $S_2$ and $S_3$, denoted $X_1$, $X_2$ and $X_3$, respectively, at time $T$ using the DM, and then compare the simulation time with that of the multi-level method. In each case we will attempt to approximate the mean populations of $S_3$ with an estimator variance of 1.0. The results from MATLAB, DM simulation are displayed in Table 8. In total, 14,500 paths were generated, taking a total of 1,070 seconds (approximately 18 minutes).

As an example of an efficient multi-level parameter set for system (32), we take $\tau_0 = 1/9$, $K = 3$ and $L = 4$, and seek an unbiased estimator for $\mathbb{E}[X_3(T)]$. This means we have six estimators, $Q_0, \ldots, Q_5^*$, which combine to produce an overall estimator, $\mathbb{Q}_u$. The multi-level method (implemented in MATLAB) gives $\mathbb{Q}_u = 3026.85 \pm 1.05$ within 222 seconds, an estimate consistent with that of the DM (see Table 8) but produced in a fifth of the time.

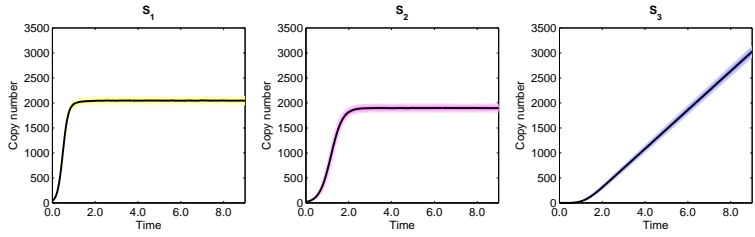

**Fig. 3** Evolution of $X^T(t) = (X_1(t), X_2(t), X_3(t))^T$ up until terminal time $T = 9$. The solid black lines show the mean species numbers and the colored bands one and two standard deviations from the mean.

| Level | $\tau_{\ell-1}$ | $\tau_\ell$ | Estimate | Sample variance | Paths | Time |
|---|---|---|---|---|---|---|
| $Q_0$ | - | $3^{-2}$ | 2951.71 | 3664.38 | 39339 | 23.7s *(152.8 s)* |
| $Q_1$ | $3^{-2}$ | $3^{-3}$ | 50.13 | 215.93 | 5910 | 16.1s *(73.0 s)* |
| $Q_2$ | $3^{-3}$ | $3^{-4}$ | 16.61 | 89.21 | 2480 | 16.1s *(62.7 s)* |
| $Q_3$ | $3^{-4}$ | $3^{-5}$ | 5.56 | 36.26 | 1138 | 18.5s *(47.0 s)* |
| $Q_4$ | $3^{-5}$ | $3^{-6}$ | 1.85 | 15.58 | 678 | 26.7s *(48.4 s)* |
| $Q_5^*$ | $3^{-7}$ | DM | 1.01 | 11.42 | 172 | 120.6s *(108.1 s)* |
| Total | | | $3026.85 \pm 1.05$ | | - | 221.8s |

**Table 9** The contribution from the estimator on each level, $Q_\ell$, in producing an unbiased overall estimator $\mathbb{Q}_u$ for $X_3$ of system (32) at $T = 9$. We have taken $\tau_0 = 1/9$, $K = 3$, and $L = 4$. In the time column, the true simulation time is shown, together with an estimated simulation time in brackets.

In order to explore our result in greater detail, we explicitly consider the contribution of each estimator $Q_\ell$ to the overall estimator $\mathbb{Q}_u$ and present our findings in Table 9. We see that 54% of simulation time is allocated to the calculation of $Q_{L+1}^*$, despite our best attempts at optimizing calculations on this final level. Unfortunately, this effort is somewhat wasted as $\mathbb{Q}_u$ is adjusted by only 0.03% with the inclusion of $Q_{L+1}^*$. Moreover, $Q_{L+1}^*$ contributes approximately 25% of the total estimator variance. With the benefit of hindsight, we can say that it may have been better to neglect the final coupling level and consider instead the corresponding biased estimator, $\mathbb{Q}_b$, with the same multi-level parameter set. In addition, we note that the true simulation times, as shown in Table 9, compare poorly with the simulation times estimated by Equation (14). We discuss this MATLAB-specific problem in Section 7.2.

The choice of the 'most efficient' parameter set for the multi-level method depends on the system to be simulated and, in general, is a non-trivial problem to solve. We now explore further the choices of $\tau_0$ and $L$ for our example system (32), and then conclude with a discussion of the final coupling level.

| Level | $\tau_{\ell-1}$ | $\tau_\ell$ | Estimate | Sample variance | Paths | Time |
|---|---|---|---|---|---|---|
| $Q_0$ | - | $3^0$ | 130.78 | 508.21 | 127,197 | 4.5s |
| $Q_1$ | $3^0$ | $3^{-1}$ | 2301.85 | 68962.36 | 539,383 | 183.9s |
| $Q_2$ | $3^{-1}$ | $3^{-2}$ | 519.10 | 67553.46 | 522,755 | 681.7s |
| $Q_3$ | $3^{-2}$ | $3^{-3}$ | 50.15 | 209.611 | 17,984 | 45.75s |
| $Q_4$ | $3^{-3}$ | $3^{-4}$ | 16.91 | 82.88 | 6,781 | 37.7s |
| $Q_5$ | $3^{-4}$ | $3^{-5}$ | 5.66 | 36.45 | 3,593 | 47.0s |
| $Q_6$ | $3^{-5}$ | $3^{-6}$ | 2.01 | 18.07 | 2,781 | 92.9s |
| $Q_7^*$ | $3^{-7}$ | DM | 1.05 | 11.38 | 477 | 154.5s |
| Total | | | $3027.15 \pm 1.12$ | | - | 1,247.8s |

**Table 10** The contribution from each level estimator $Q_\ell$ in producing an unbiased overall estimator $\mathbb{Q}_u$ for $X_3$ at $T = 9$. We have taken $\tau_0 = 1$, $K = 3$, and $L = 4$.

| Guess | Estimates | |
|---|---|---|
| $\tau_0^{(1)} = 3$ | $\sqrt{k_0^{(1)}} = 4.8609$ | $\sqrt{k_1^{(1)}} = 9.1735$ |
| $\tau_0^{(2,1)} = 1$ | $\sqrt{k_0^{(2,1)}} = 0.1701$ | $\sqrt{k_1^{(2,1)}} = 6.0573$ |
| $\tau_0^{(2,2)} = 9$ | $\sqrt{k_0^{(2,2)}} = 0.0028$ | $\sqrt{k_1^{(2,2)}} = 5.1600$ |
| $\tau_0^{(3)} = 1/3$ | $\sqrt{k_0^{(3)}} = 5.8497$ | $\sqrt{k_1^{(3)}} = 10.0359$ |
| $\tau_0^{(4)} = 1/9$ | $\sqrt{k_0^{(4)}} = 1.7699$ | $\sqrt{k_1^{(4)}} = 0.8384$ |
| $\tau_0^{(5)} = 1/27$ | $\sqrt{k_0^{(5)}} = 3.0284$ | N/A |

**Table 11** Details of the cost measure for each potential ensemble of estimators for (32) with different choices of $\tau_0$, the time step on the base level.

5.1 Choice of the base level

In Table 10 we demonstrate that an inappropriate choice of $\tau_0$ in the multi-level method can lead to a dramatically increased simulation time (1,247.8 seconds, compared with 221.8 seconds). Whilst this choice of $\tau_0$ results in a reasonable estimate of $\mathbb{E}[X_3(T)]$, the CPU time required is greater than that of our DM. Looking in detail at the CPU time required for each level, we see that the estimator $Q_0$ is calculated within 4.5 seconds, but the estimator $Q_1$ takes 183.9 seconds to compute and $Q_2$ takes 681.7 seconds. The base level is too inaccurate to capture the salient details of the system, and consequently these must be restored with the subsequent estimators.

In order to avoid this situation, we need to choose $\tau_0$ in an intelligent manner. To do this we follow the method suggested in Section 4.1. As a first guess, take $\tau_0^{(1)} = 3$. Using the results displayed in Table 11 we refine our choice of $\tau_0$:

1. We let $\tau_0^{(2,1)} = 1$, $\tau_0^{(2,2)} = 9$ be the candidate refinements of $\tau_0$. We have

$$\sqrt{k_0^{(1)}} + \sqrt{k_1^{(1)}} > \sqrt{k_0^{(2,1)}},$$
$$\sqrt{k_0^{(2,2)}} + \sqrt{k_1^{(2,2)}} > \sqrt{k_0^{(1)}},$$

and so we accept $\tau_0^{(2,1)}$ and reject $\tau_0^{(2,2)}$. We therefore set $\tau_0^{(2)} = \tau_0^{(2,1)}$.

| $\tau_0$ | $\tau_L$ | L | K | Estimate | Duration |
|---|---|---|---|---|---|
| $3^0$ | $3^{-6}$ | 6 | 3 | $3027.15 \pm 1.12$ | 1247.83 s |
| $3^{-1}$ | $3^{-6}$ | 5 | 3 | $3027.38 \pm 1.11$ | 1223.35 s |
| $3^{-2}$ | $3^{-6}$ | 4 | 3 | $3026.46 \pm 1.03$ | 208.34 s |
| $3^{-3}$ | $3^{-6}$ | 3 | 3 | $3027.03 \pm 1.06$ | 224.59 s |

**Table 12** Various estimates of $\mathbb{E}[X_3(T)]$ for system (32) generated using the unbiased multi-level method with different base level time steps. Note that $\tau_L$ now refers to the overall accuracy of the biased estimator, before the final estimator $Q^*_{L+1}$ is included to produce an unbiased estimator.

2. We let $\tau_0^{(3)} = \tau_0^{(2)}/3 = 1/3$. As

$$\sqrt{k_0^{(2)}} + \sqrt{k_1^{(2)}} > \sqrt{k_0^{(3)}},$$

we accept $\tau_0^{(3)}$ as an improvement.

3. We let $\tau_0^{(4)} = \tau_0^{(3)}/3 = 1/9$. As

$$\sqrt{k_0^{(3)}} + \sqrt{k_1^{(3)}} > \sqrt{k_0^{(4)}},$$

we accept $\tau_0^{(4)}$ as an improvement.

4. We let $\tau_0^{(5)} = 1/27$. However,

$$\sqrt{k_0^{(4)}} + \sqrt{k_1^{(4)}} < \sqrt{k_0^{(5)}},$$

and so we reject $\tau_0^{(5)}$, and conclude that a good choice is $\tau_0 = 1/9$.

We therefore, under the restriction of $K = 3$, fix $\tau_0 = 1/9$ and proceed with the rest of the method.

In Table 12 we demonstrate the performance of a range of alternative base level time steps with a corresponding change in number of levels so that the biased estimators that would be produced (before the final, exact, coupling) always have the same level of bias. $\tau_0 = 1/9$ is, as predicted by our method for choosing the base level time step, the most efficient choice for the example system (32)

5.2 Choice of the final level

As mentioned in the previous section, it may be prudent to use the MSE as an accuracy metric. We recall that

$$\text{MSE} = \text{Var}(\mathbb{Q}) + \{\text{Bias}(\mathbb{Q}, \mathbb{E}[X_3(T)])\}^2, \qquad (33)$$

and we seek to bound this by some $\varepsilon^2$. We choose $\lambda \in (0,1)$ and then aim to bound the estimator variance of $\mathbb{Q}$ by $(1-\lambda)\varepsilon^2$, and the bias by $\sqrt{\lambda}\varepsilon^2$. For our sample problem (32) we have taken $\tau_0 = 1/9$, $K = 3$, and sought to estimate $\mathbb{E}[X_3(T)]$ subject to a maximum MSE of 1.0. In Table 13 we present a range of results for different choices of $\lambda$. Our answers, unsurprisingly, compare very favorably with the exact estimates given in Table 8. Moreover, our algorithm ensures that the required MSE has been attained.

| $\lambda$ | Estimate | Levels | Time (sec) |
|---|---|---|---|
| 0.95 | $3026.18 \pm 0.44$ | 5 | 103.07 |
| 0.75 | $3026.29 \pm 0.98$ | 6 | 154.05 |
| 0.50 | $3026.67 \pm 1.38$ | 6 | 97.27 |
| 0.25 | $3026.67 \pm 1.70$ | 6 | 72.92 |
| 0.05 | $3027.52 \pm 1.91$ | 7 | 397.82 |

**Table 13** Details of the simulation times for each potential ensemble of estimators for (32) with different choices of $\lambda$, the proportion of the total MSE assigned to the bias. The value of $\lambda$ has been fixed, and the procedure of Section 4.3 followed.

This approach is simpler to implement than the unbiased method and can be easily automated. By contrast, the unbiased approach cannot be easily automated, and the decision as to how many levels to incorporate before the *final* coupling is implemented is far from obvious. Future work could focus on the optimal choice of $\lambda$, or indeed, an adaptive choice of $\lambda$ as the simulation progresses. This could potentially take full account of the fact that the number of estimators incorporated will be necessarily discrete.

## 6 A third example system

In this section, we consider a third example system. Our simulations will be conducted in `C++`. This model describes the mitogen-activated protein kinase (MAPK) cascade, which is involved in a wide variety of signalling processes that govern transitions within the phenotype of a cell, and has previously been used as a test case for stochastic simulation algorithms (MacNamara et al, 2008). This model comprises ten coupled Michaelis-Menten schemes (Huang and Ferrell, 1996), and has $N = 22$ species and $M = 30$ reactions. A Michaelis-Menten scheme is constructed as follows (MacNamara et al, 2008): there are four species and three reaction channels within the scheme. The species are substrate ("S"), enzyme ("E"), complex ("ES") and product ("P") particles. The reaction channels are as follows:

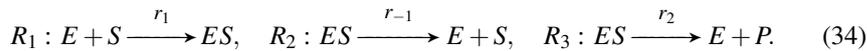

$$R_1 : E + S \xrightarrow{r_1} ES, \quad R_2 : ES \xrightarrow{r_{-1}} E + S, \quad R_3 : ES \xrightarrow{r_2} E + P. \quad (34)$$

A quasi-steady state assumption can then be applied to reduce the computational complexity associated with simulating the system. This reduces the scheme to two species: substrate ("S") and product ("P") particles. The three reaction channels described by (34) are reduced into a single reaction channel, which is given as

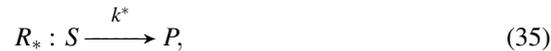

$$R_* : S \xrightarrow{k^*} P, \quad (35)$$

where the propensity function follows Michaelis-Menten kinetics, which are given by

$$k^* = \frac{k_2 E_0 S}{S + \frac{r_{-1} + r_2}{r_1}},$$

where $E_0$ represents the initial enzyme population. As explained, the MAPK cascade comprises ten coupled Michaelis-Menten schemes: we provide a diagrammatic representation in Figure 4. We will now simulate the model using the quasi-steady state assumption. This means we will simulate a model comprising ten reaction channels. The propensity value of each reaction channel is given by the Michaelis-Menten kinetic formula detailed in equation (35). The substrate, enzyme and product particles for each channel are as shown in Figure 4. The reaction channels are therefore as follows:

$$\begin{aligned}
R_1 &: KKK \xrightarrow{k_1} KKK\text{-}P, & R_2 &: KKK\text{-}P \xrightarrow{k_2} KKK, \\
R_3 &: KK \xrightarrow{k_3} KK\text{-}P, & R_4 &: KK\text{-}P \xrightarrow{k_4} KK, \\
R_5 &: KK\text{-}P \xrightarrow{k_5} KK\text{-}PP, & R_6 &: KK\text{-}PP \xrightarrow{k_6} KK\text{-}P, \\
R_7 &: K \xrightarrow{k_7} K\text{-}P, & R_8 &: K\text{-}P \xrightarrow{k_8} K, \\
R_9 &: K\text{-}P \xrightarrow{k_9} K\text{-}PP, & R_{10} &: K\text{-}PP \xrightarrow{k_{10}} K\text{-}P,
\end{aligned} \qquad (36)$$

where the $k_j$ are functions of the form of Equation (35). We will estimate the mean MAPK population (indicated by "K-P" in System (36) and Figure 4) at a terminal time $T$. We use suitable initial conditions to simulate the model until a terminal time $T = 250$. The initial conditions are detailed in Table 14. We now provide the model parameters. Each Michaelis-Menten reaction is of the form $R_j : X \xrightarrow{k_j} Y$, and the function $k_j$ is expressed as $k_j = \alpha_j \cdot X / (X + \beta_j)$. For each reaction $R_j$, the initial enzyme populations give $\alpha_j$ and $\beta_j$ their values. We use the following values: $\alpha_1 = 2.5$, $\alpha_2 = 0.25$, $\alpha_3 = 0.025$, $\alpha_4 = 0.75$, $\alpha_5 = 0.025$, $\alpha_6 = 0.75$, $\alpha_7 = 0.025$, $\alpha_8 = 0.5$, $\alpha_9 = 0.025$, and $\alpha_{10} = 0.5$. We also have $\beta_1 = 10$, $\beta_2 = 8$, $\beta_3 = 15$, $\beta_4 = 15$, $\beta_5 = 15$, $\beta_6 = 15$, $\beta_7 = 15$, $\beta_8 = 15$, $\beta_9 = 15$, and $\beta_{10} = 15$. If the DM is used, it will take approximately 7,300 seconds to estimate the mean MAPK population at time $T$ to a suitable level of accuracy in C++ (approximately two hours). In Table 15 we show a multi-level configuration which estimates the mean MAPK population as $2682.87 \pm 0.10$; this calculation in C++ takes approximately 1,376 seconds. This demonstrates that with even a relatively complicated reaction network, a five-fold reduction in simulation time can be achieved with the multi-level method.

## 7 Discussion and outlook

This final section discusses some remaining challenges in implementing the multi-level method. Under specific circumstances, these challenges might make it difficult to implement the multi-level method. We discuss one particular difficulty which arises where two simultaneously-generated paths grow far apart over time; a number of potential remedies are then discussed. We proceed to discuss a range of implementation issues.

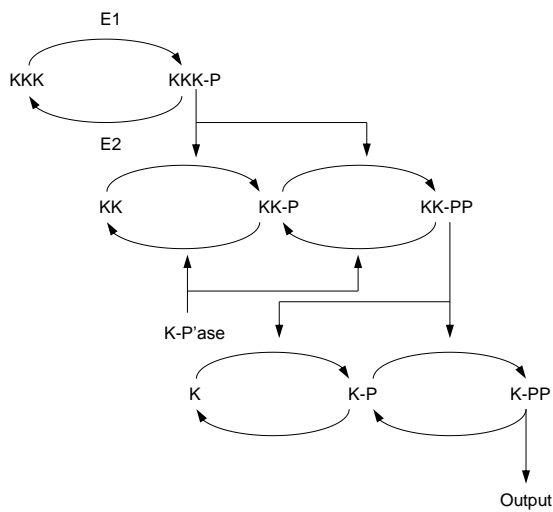

**Fig. 4** A diagrammatic representation of the MAPK cascade. The text refers to chemical species; whilst the arrows represent Michaelis-Menten schemes. The arrow points from the substrate towards the product; the species on top of the arc indicates the enzyme. This diagram has been adapted from Huang and Ferrell (1996).

| Species | Initial Value | Species | Initial Value |
|---|---|---|---|
| *KKK* | 90 | *KKK-P* | 10 |
| *KK* | 280 | *KK-P* | 10 |
| *KK-PP* | 10 | *K* | 280 |
| *K-P* | 10 | *K-PP* | 10 |

**Table 14** This table provides the initial values for the MAPK cascade model detailed in (36).

| Level | $\tau_{\ell-1}$ | $\tau_\ell$ | Estimate | Sample variance | Paths | Time |
|---|---|---|---|---|---|---|
| $Q_0$ | - | $4^{-2}$ | 2331.79 | 15846.10 | $1.24 \cdot 10^6$ | 681.2 s |
| $Q_1$ | $4^{-2}$ | $4^{-3}$ | 275.99 | 537.44 | $1.15 \cdot 10^6$ | 243.9 s |
| $Q_2$ | $4^{-3}$ | $4^{-4}$ | 57.59 | 166.21 | 424012 | 211.8 s |
| $Q_3$ | $4^{-4}$ | $4^{-5}$ | 13.21 | 31.75 | 121385 | 146.7 s |
| $Q_4^*$ | $4^{-6}$ | DM | 4.29 | 11.19 | 58235 | 100.7 s |
| **Total** | | | **2682.87 ± 0.10** | | - | **1,384.3s** |

**Table 15** The contribution from each level estimator $Q_\ell$ in producing an unbiased overall estimator for the mean MAPK population at time $T = 250$. We have taken $\tau_0 = 1/16$, $K = 4$, and $L = 3$.

| Estimator | 0.01 | 0.1 | 1 | 5 | 50 | 95 | 99 | 99.9 | 99.99 |
|---|---|---|---|---|---|---|---|---|---|
| $Q_3$ | -360.5 | -311 | 10 | 21 | 40 | 63 | 74 | 87 | 198.5 |
| $Q_4$ | -379 | -256.5 | 3 | 6 | 13 | 22 | 27 | 32 | 43 |
| $Q_5$ | -349.5 | -58.5 | 0 | 1 | 4 | 9 | 11 | 13 | 15 |

**Table 16** Sample values for various percentiles in the distributions used to estimate $Q_3$, $Q_4$ and $Q_5$ for the gene expression system (7) using $\tau_0 = 1/9$ and $K = 3$. 100 000 data points have been used for each estimator.

### 7.1 Catastrophic decoupling

Consider the contribution of each term in $\mathbb{Q} = Q_0 + Q_1 + \cdots + Q_L(+Q_{L+1}^*)$ to the multi-level estimator. In the course of our exploration of the multi-level method, we have noticed that occasionally sample paths on one level undergo what we will call a 'catastrophic decoupling' so that species populations in a pair of sample paths become very different from one another. This can have a dramatic effect on the variance of the estimator on that level, and hence on the results of the multi-level method. For example, if such a sample path is generated in the course of estimating the $n_\ell$, the optimization algorithm of Section 3.2 then suggests that a huge number of samples are needed on that level. This slows the multi-level calculation, and the result is often that the actual variance of the estimator is much lower than the target variance. On the other hand, if we do not see a catastrophic decoupling during estimation of the $n_\ell$, but one or more occur during generation of sample paths for the $Q_\ell$, the target variance for the estimator $\mathbb{Q}$ is not achieved.

We now give an example of a catastrophic decoupling event. For the example gene expression system (7), we take $\tau_0 = 1/9$ and $K = 3$. In Table 16 we show percentile data for distributions of $Q_3$, $Q_4$ and $Q_5$. It is clear that the sample values contributing to $Q_3, Q_4$ and $Q_5$ all possess extreme tails to their distributions as a result of one or more catastrophic decoupling events. For example, over 90% of the sample values for $Q_5$ lie in the interval $[1,9]$, but approximately 1 in 1150 sample paths provide sample values of $-100$ or less. This makes catastrophic decoupling events appear deceptively unlikely; however, if 100 sample paths are generated, there is an 8% chance that such an event is encountered. If 1000 sample paths are performed, this rises to approximately 58%.

We now explain the cause of this problem, and then discuss its consequences. In effect, a decoupling is possible each time a new mRNA molecule (*M*) is introduced into the system. The coupling technique ensures it is introduced into *both* the coarse and fine systems. In the fine system, the decay process of this mRNA starts immediately. However, in the coarse system this is not always the case: this is because decay of the mRNA cannot take place in the coarse system until the reaction propensities are updated. Hence, during this interim period, it is possible for the new mRNA particle to decay in the fine system but not in the coarse system.

It is clear that the scaling of the system is then what causes problems with the variance. At time $T = 1$, there are approximately 24 mRNA molecules, compared with over $3,000$ protein molecules. If the decoupling in mRNA species counts occurs at an early time, the extra mRNA molecule in the coarse system leads to increased protein generation which, in turn, leads to increased dimer generation. This difference in generation rates remains until the mRNA populations converge again (if at all). As the dimer population is monotonically increasing, the population difference is 'locked in' for all subsequent times, and so the difference in sample values of $X_3(T)$ is large.

| Sample | Sample mean | Sample variance |
|---|---|---|
| No decouplings | 4.33 | 4.82 |
| Single decoupling | 4.11 | 90.41 |

**Table 17** Statistics demonstrating two different outcomes when 1000 sample paths are produced to estimate $Q_5$ for system (7). The system without decouplings has all its data points in [-1; 13], whilst the other system has a single decoupling, value -288. The rest of the data is contained in [-2; 12].

| Estimator | Mean | Variance | Kurtosis |
|---|---|---|---|
| $Q_3$ | 39.33 | 646.15 | 108.41 |
| $Q_4$ | 13.07 | 197.23 | 438.00 |
| $Q_5$ | 4.34 | 65.62 | 1257.80 |

**Table 18** Statistics describing the samples for $Q_3, Q_4$ and $Q_5$ for system (7) using $\tau_0 = 1/9$ and $K = 3$. 100 000 data points have been used for each estimator.

As previously noted, when a simulation exhibiting a catastrophic decoupling is incorporated into a Monte Carlo estimator, it has a substantial effect on the estimator variance. However, it is often the case that the estimator itself is relatively unaffected. In Table 17 we show the dramatic effect that just a single decoupling has on the sample variance for one level in the gene expression system (7), without having an overwhelming effect on the mean estimate.

Here we will not present a concrete method which avoids these catastrophic decoupling; we suggest this as an area for future work. A first possibility lies in the use of common inhomogeneous Poisson processes for each of the coarse and fine sample paths. In Table 18 we provide sample means, variances and kurtoses for the gene expression system. This demonstrates that higher estimators are associated with substantially higher kurtoses: a second possibility could involve the use of the kurtosis to detect the presence of a decoupling (Bayer et al, 2014).

### 7.2 Implementation challenges

With this work, we provide code for the multi-level method written in both `MATLAB` and `C++`, and now we discuss implementation using these two platforms. When sample paths are produced with `MATLAB`, the code must be 'vectorized' to ensure a high level of efficiency. To highlight the benefits of this approach, we note that our DM results for the system (7) shown in Table 2 took around six hours to produce, whereas a non-vectorized DM code could require 162 hours (nearly a week).

One of the highly reaction system-dependent components of the multi-level technique is the optimization algorithm used to choose $n_\ell$, the number of samples required on each level. The algorithm should, of course, aim to provide the optimal number of simulations required on each level; we have followed Anderson and Higham (2012) in generating 100 initial simulations to guide this choice. However, the results in Table 9 show that the actual simulation time is different to that predicted by the optimization algorithm. In `MATLAB` this is largely because code vectorization means that, for example, the CPU time per path when generating, say, 100 sample paths is different (and usually greater) than the CPU time per path when generating 1000 paths. As such, the optimization algorithm does not work as well for `MATLAB` vectorized code as it does for code implemented using `C++`, as it over-estimates the simulation time: especially where very many simulations are required.

7.3 Higher order estimators

In this manuscript we have focussed on estimating only the mean and variance of a population within a chemical reaction network. The method can be naturally extended to estimate other summary statistics, for example, the signal-to-noise ratio. One particularly interesting challenge is to deal with systems which comprise multiple favourable states, such as the Schlogl system (Vellela and Qian, 2009). In this case, the multi-level method can be used to estimate the $r$-th moments of the copy number, $\mu_r = \mathbb{E}[X^r]$, for $r = 1, \ldots, C$. These moments can then be used to construct an approximate probability distribution for the copy number by using the Method of Moments (Kavehrad and Joseph, 1986). We leave the details to a future work.

7.4 Summary

The multi-level method provides the potential for great savings to be made in the world of stochastic simulation of chemical systems. Although there are many intricacies associated with the method, many of them software- and system-dependent, the benefits of using multi-level approaches are enormous, and open up the range of problems that can be fully explored using stochastic simulation. We have introduced a number of novel enhancements to the multi-level method which, we feel, make it easier to understand and implement, as well more computationally efficient.